\documentclass[preprint2]{aastex}
\usepackage{psfig}

\newcommand{\minusone}{$^{-1}$} 
\newcommand{\LOH}{$L_{OH}$ }
\newcommand{\LFIR}{$L_{FIR}$ }

\righthead{OH Megamasers II:  Further Results}
\lefthead{Darling \& Giovanelli}

\begin{document}
\title{A Search for OH Megamasers at $z > 0.1$. \ \ II.\  Further Results}
\author{Jeremy Darling \& Riccardo Giovanelli}
\affil{Department of Astronomy and National Astronomy and Ionosphere Center, 
Cornell University}
\authoraddr{Department of Astronomy, Cornell University, Ithaca,  NY  14853;
	darling@astrosun.tn.cornell.edu; riccardo@astrosun.tn.cornell.edu}

\begin{abstract}
We present current results of an ongoing survey for OH megamasers 
in luminous infrared galaxies at the Arecibo Observatory.  The survey is 
now two-thirds complete, and has resulted in the discovery of 35 new 
OH megamasers at $z>0.1$, 24 of which are presented in this paper.  
We discuss the properties of each source in detail, including 
an exhaustive survey of the literature.  We also place upper limits 
on the OH emission from 107 nondetections and list their IR, radio, and
optical properties.  
The survey detection rate is 1 OH megamaser for every 6 candidates overall,
but is a strong function of the far IR luminosity of candidates and may depend
on merger stage or on the central engine responsible for the IR luminosity
in the merging galaxy pair.  We also report the detection of {\it IRAS} 
12032+1707, a new OH gigamaser.
\end{abstract}

\section{Introduction}

The first paper in this series presents the motivation, goals,
and preliminary results of a survey for OH megamasers (hereafter OHMs)
at the Arecibo Observatory\footnote{The Arecibo
Observatory is part of the National Astronomy and Ionosphere Center, which 
is operated by Cornell University under a cooperative agreement with the
National Science Foundation.}
(Darling \& Giovanelli 2000; hereafter Paper I).  
To recap the survey, we use the Point Source Catalog redshift survey 
(PSCz; Saunders et al. 2000)
to select OHM candidates which lie within the Arecibo declination range 
$0^\circ<\delta<37^\circ$ and have $z>0.1$.  The requirement for candidates
to have a measured redshift $z>0.1$ and an {\it IRAS} detection at 60 $\mu$m 
guarantees that they are (ultra)luminous
IR galaxies ([U]LIRGs).  The survey is designed to double the sample of OHMs
to roughly 100, and increase the $z>0.1$ sample by a factor of seven.  
The Arecibo OHM survey will provide a uniform flux-limited sample which 
can be tied to the low-redshift FIR luminosity function (see Paper I).  
One can subsequently
perform deep searches for OH line emission at various redshifts and effectively
measure the evolution of the FIR luminosity function with redshift.  
Since ULIRGs are major mergers \citep{cle96}, one can also
measure the number evolution of galaxies by measuring
the number density of OHMs at various redshifts \citep{briggs98}.
A measure of the number density evolution of galaxies breaks the degeneracy
between luminosity evolution and number density evolution seen in the 
luminosity function evolution derived from redshift surveys 
(eg - Le F\'{e}vre et al. 2000; Lilly et al. 1995; Ellis et al. 1996) 
and provides strong 
constraints on hierarchical models of galaxy formation and evolution
(eg - Abraham 1999; Governato 1999).  The OHM sample will also
provide new perspective on OH megamaser environments, lifetimes, 
engines, and structure.  The $z>0.1$ sample is distant enough that many
masing regions will have sufficiently small angular scales to show 
interstellar scintillation \citep{wal98}.  
Indeed, several OHMs have been observed to
vary in time, and a comprehensive study of OHM variability is underway.  
Many of the OHM hosts have optical spectral classification of their nuclei
available in the literature, 
but those which do not are under investigation at the Palomar 5m
telescope.  Finally, as in any survey, the Arecibo OHM survey
is likely to produce unexpected results or detections which cannot
be predicted {\it a priori}.  Several of the new detections in this paper
and Paper I have unexpected properties (see \S \ref{detect}), and  
more surprises are likely before the survey is complete. 

This paper is primarily dedicated to releasing the new OH megamaser
detections for rapid community access.  Paper I presented
11 OHMs, one OH absorber, and 53 nondetections, accounting for $22\%$
of the survey.  This paper presents 24 new OHMs and 107 nondetections, 
accounting for $44\%$ of the total survey.  We discuss the observations,
data reduction, and newly implemented RFI mitigation techniques in Section 
\ref{obs}, the new detections as well as the non-detections
in Section \ref{results}, and the results to date and the 
plan for the remainder of the survey in Section \ref{discussion}.

This paper parameterizes the Hubble constant as $H_\circ = 75\ h_{75}$
km s\minusone\ Mpc\minusone, assumes $q_\circ = 0$, and 
uses $D_L = (v_{CMB} / H_\circ)(1+0.5z_{CMB})$ to 
compute luminosity distances from $v_{CMB}$, the radial velocity of the 
source in the cosmic microwave
background (CMB) rest-frame.  Line luminosities are always computed
under the assumption of isotropic emission.

\section{Observations and Data Reduction}\label{obs}

Observations at Arecibo with the L-band receiver were performed in an 
identical manner to earlier survey work 
by nodding on- and off-source, each for a 4-minute total integration, followed
by firing a noise diode (see Paper I).  
Spectra were recorded in 1-second intervals (a significantly faster dump
rate than usual for spectroscopy work at Arecibo)
to facilitate radio frequency interference (RFI) 
flagging and excision in the time-frequency domain.
Data was recorded with 9-level sampling in 2 polarizations
of 1024 channels each, spanning a 25 MHz bandpass centered on redshifted
1666.3804 MHz (the mean of the 1667.359 and 1665.4018 MHz OH lines).  

\begin{figure}
\plotone{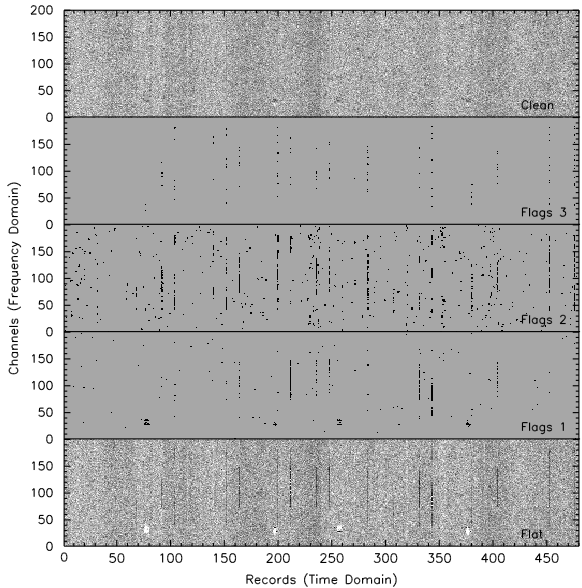}
\caption{Frequency-time slices of an unusually dense radio frequency 
interference (RFI) environment showing
several stages of the RFI flagging and cleaning procedure.
Each channel spans 24.4 kHz, and a 200 channel slice is shown in the
frequency domain.  The horizontal scale
represents a time series of 1 second records;  the first 240 records
are the on-source position and the last 240 records are the off-source 
position.  Panels from bottom
to top illustrate: a flattened ``dirty'' image in which the rate of 
occurrence and the spectral extent of RFI is emphasized by the dark 
pixels ({\it Flat}); 
the flags array produced in the first time-domain filter ({\it Flags 1}); 
the flags array produced in the second
time-domain filter after spectral boxcar averaging ({\it Flags 2}); 
the flags array 
produced from near neighbors ({\it Flags 3}); 
and the final cleaned image ({\it Clean}).  
The greyscale is normalized such that the median grey color (such as
seen in the background of the {\it Flags} panels) represents zero, 
dark pixels are positive values, and light pixels are negative.  Flagged
channels in the {\it Flags} arrays are set to unity (black), whereas 
flagged channels in the cleaned image are set to zero (they are not in fact
included in the final time averaged spectrum).
\label{rfi}}
\end{figure}

All data reduction and analysis was performed with the AIPS++
\footnote{The AIPS++ (Astronomical Information Processing System) is a 
product of the AIPS++ Consortium. AIPS++ is freely available for use under 
the Gnu Public License. Further information may be obtained from 
http://aips2.nrao.edu}
software package
using home-grown routines for single dish reduction.  We included 
in the reduction pipeline RFI flagging routines designed to identify two
types of RFI observed at Arecibo:  strong features more
than $3\sigma$ above the time-domain noise and spectrally broad, 
low-level RFI which is time-variable.  Figure \ref{rfi} illustrates
with time-frequency array images some of the RFI flagging steps outlined
below.  The reduction pipeline and RFI flagging procedure follow the
steps: (1) form an array in time and frequency from a 4-minute 
on-off pair (1024 channels $\times$ 480 1-second records); (2) flatten the
array ``image'' by subtracting the time-averaged or median (both are examined)
bandpass shape 
(average/median of 240 records; see Figure \ref{rfi},  {\it Flat}); 
(3) compute an RMS noise spectrum (RMS across the time domain)
and fit this spectrum with a high-order polynomial; (4) compare
the smooth polynomial noise spectrum to individual records and flag 
$3\sigma$ channels ({\it Flags 1} in Figure \ref{rfi}); 
(5) perform a boxcar spectral
average over 5 channels for each record and repeat the RMS noise spectrum 
fitting and channel flagging ({\it Flags 2} in Figure \ref{rfi}); (6) flag 
channels which have at least 8 flagged neighbors in a 10 channel box in the
spectral regime ({\it Flags 3} in Figure \ref{rfi});  
(7) set all flagged channels to zero in the flattened data to obtain a clean 
image ({\it Clean} in Figure \ref{rfi});
(8) form an average on- and off-source spectrum from the original raw data, 
rejecting flagged channels and keeping track of the number of records 
averaged in each channel; (9) perform the usual (on$-$off)/off conversion
to obtain an intensity scale expressed as a percentage of the system 
temperature, and use VLA calibrators and noise diodes
to convert to flux density units (mJy); (10) average multiple on-off pairs
and polarizations weighted by the number of records used in each channel, 
keeping track of the effective noise present in each channel; (11) fit and 
subtract baselines and hanning smooth to obtain the final spectra.
The frequency resolution after hanning smoothing is 49 kHz (10 km s\minusone\
at $z=0.1$), and the uncertainty in the absolute flux scale is $8\%$.  
The RFI flagging procedure was tested on both synthetic and real data with
the result that time-constant signals are unaffected by the process to within
roundoff errors introduced in the processing, and that a reliable estimate of
the weight, or effective integration time, on a single channel is the 
total number of records used in forming the time average, excluding flagged 
records.  Each spectral channel in a 
calibrated time-averaged spectrum may have a different effective integration
time and hence a different effective intrinsic radiometer noise level.
We include a normalized weights spectrum with each OHM spectrum in Figure 
\ref{spectra}.  Depressions in the weights spectra indicate the presence of 
RFI, which may or
may not have been completely flagged and removed from the final spectra.  
Spectral channels with low weights should thus be treated with skepticism.

The techniques described here were {\it not} performed on the data presented
in Paper I.  We applied these data reduction steps to several of the Paper I
spectra and compared the results to the published spectra.  
Spectra are not significantly changed
except in cases of significant RFI (the usual averaging process tends to 
mitigate low-level RFI anyway),
although the ANALYZ reduction process used for the Paper I spectra 
tends to produce lower overall RMS noise 
levels.  We attribute this to the automatic application of hanning smoothing to
each 6-second record in ANALYZ, rather than applying hanning smoothing to 
the final spectrum as we do in AIPS++.  Electronic versions of all spectra 
processed following the methods described in this paper, for the sources in 
both papers, are available upon request.

\section{Results}\label{results}

\subsection{Nondetections}\label{nondetect}

Tables \ref{nondetectFIR} and \ref{nondetectOH} list respectively 
the optical/FIR and radio properties of 107 OH non-detections.
Note that IRAS 03477+2611 also appears in Paper I;  reobservation
produced a stronger OH flux limit on this candidate, as listed in 
Table \ref{nondetectOH} of this paper.
Table \ref{nondetectFIR} lists the optical redshifts
and FIR properties of the non-detections in the following format.
Column (1):  {\it IRAS} Faint Source Catalog (FSC) name.  
Columns (2) and (3):  Source coordinates (epoch B1950.0) 
from the FSC, or the Point Source Catalog (PSC) if unavailable in the FSC.  
Columns (4), (5) and (6):  Heliocentric optical redshift, 
reference, and corresponding
velocity.  Uncertainties in velocities are listed whenever they are 
available.  
Column (7):  Cosmic microwave background rest-frame velocity.  This is
computed from the heliocentric velocity using the solar motion with respect
to the CMB measured by \citet{lin96}:  $v_\odot = 368.7 \pm 2.5$ km s\minusone\
towards $(l,b) = (264\fdg31 \pm 0\fdg16 , 48\fdg05 \pm 0\fdg09)$.
Column (8):  Luminosity distance computed from $v_{CMB}$ via 
$D_L = (v_{CMB} / H_\circ)(1+0.5z_{CMB})$, assuming $q_\circ = 0$.  
Columns  (9) and (10):  {\it IRAS} 60 and 100 $\mu$m flux 
densities in Jy.  FSC flux densities are listed whenever they are 
available.  Otherwise, PSC flux densities
are used.  Uncertainties refer to the last digits of each measure, and upper 
limits on 100 $\mu$m flux densities are indicated by a ``less-than'' symbol. 
Column (11):  The logarithm of the far-infrared luminosity in units
of $h_{75}^{-2}L_\odot$.  
\LFIR is computed following the prescription of \citet{ful89}:  
\LFIR$ = 3.96\times 10^5 D_L^2 (2.58 f_{60} + f_{100})$, 
where $f_{60}$ and $f_{100}$ are the 60 and 100 
$\mu$m flux densities expressed in Jy, 
$D_L$ is in $h_{75}^{-1}$Mpc, and \LFIR is in units of $h_{75}^{-2}L_\odot$.  
If $f_{100}$ is only available as an upper limit, the permitted range
of \LFIR is listed.  The lower bound on \LFIR is computed for $f_{100}=0$ mJy,
and the upper bound is computed with $f_{100}$ set equal to its upper limit.
The uncertainties in $D_L$ and in the {\it IRAS} flux densities 
typically produce an uncertainty in $\log L_{FIR}$ of $0.03$.  

Table \ref{nondetectOH} lists the 1.4 GHz 
flux density and the limits on OH emission of the non-detections in 
the following format:
Column (1):  {\it IRAS} FSC name, as in Table \ref{nondetectFIR}.
Column (2):  Heliocentric optical redshift, as in Table \ref{nondetectFIR}.
Column (3):  $\log$ \LFIR, as in Table \ref{nondetectFIR}.
Column (4):  Predicted isotropic OH line luminosity, 
$\log L_{OH}^{pred}$,
based on the Malmquist bias-corrected $L_{OH}$-\LFIR relation 
determined by \citet{kan96} for 49 OHMs:  
$\log L_{OH} = (1.38\pm0.14) \log L_{FIR} - (14.02\pm 1.66)$ 
(see Paper I).
Column (5):  Upper limit on the isotropic OH line luminosity, 
$\log L_{OH}^{max}$.  
The upper limits on \LOH are computed from the RMS noise of the non-detection
spectrum assuming a ``boxcar'' line profile of rest frame width 
$\Delta v = 150$ km s\minusone\ and height 1.5 $\sigma$: 
$\log L_{OH}^{max} = \log (4 \pi D_L^2\ 1.5 \sigma [\Delta v/c] 
[\nu_\circ /(1+z)])$.   The assumed rest frame width $\Delta v = 150$ km
s\minusone\ is the average FWHM of the 1667 MHz line of the known OHM sample.
Column (6):  On-source integration time, in minutes.  
Column (7):  RMS noise values in flat regions of the 
non-detection baselines, in mJy, after spectra were hanning smoothed 
to a spectral resolution of 49 kHz.
Column (8):  1.4 GHz continuum fluxes, from the NRAO 
VLA Sky Survey (NVSS; Condon et al. 1998).  If no continuum source lies within 
30$\arcsec$ of the {\it IRAS} coordinates, an upper limit of 5.0 mJy 
is listed.  
Column (9):  Optical spectroscopic classification, if available.  
Codes used are:  ``S2'' = Seyfert type 2;
``S1'' = Seyfert type 1;  ``A'' = active nucleus;  ``C'' = composite
active and starburst nucleus;  ``H'' = \ion{H}{2} region (starburst);  
and ``L'' = low-ionization emission region (LINER).  References
for the classifications are listed in parentheses and included at the 
bottom of the Table.
Column (10):  Source notes, listed at the bottom of the Table.

We can predict the expected isotropic OH line luminosity, $L_{OH}^{pred}$, 
for the OHM candidates based on the 
$L_{OH}$-\LFIR relation determined by Kandalian (1996; see Paper I) 
and compare this figure to upper limits on the OH emission derived
from observations, $L_{OH}^{max}$, for a rough measure of the 
confidence of the non-detections.  Note, however, 
that the scatter in the $L_{OH}$--$L_{FIR}$ relation is quite large:  roughly 
half an order of magnitude in \LFIR and one order of magnitude in \LOH (see
Kandalian 1996).  Among the non-detections, 18 out of 107 galaxies have 
$L_{OH}^{pred} < L_{OH}^{max}$, indicating that longer integration times 
are needed to unambiguously confirm these non-detections, and
13 out of 107 candidates have $L_{OH}^{max}$ within the range of 
$L_{OH}^{pred}$ set by an upper limit on $f_{100}$.  
Integration times were a compromise between efficient use of telescope
time and the requirement for a meaningful upper limit on \LOH for
non-detections.
Given the scatter of identified OHMs about the $L_{OH}$--$L_{FIR}$
relation, we estimate that there are less than 8 additional OHMs
among the non-detections, but this estimate relies on uncertain
statistics of small numbers.  A thorough analysis of
completeness will be performed once the survey is complete.

\subsection{New OH Megamaser Detections}\label{detect}

Tables \ref{detectFIR} and \ref{detectOH} list respectively the 
optical/FIR and radio properties of the 24 new OHM detections.  
Spectra of the 24 OHMs appear
in Figure \ref{spectra}.  
The column headings of Table \ref{detectFIR} are identical to those of Table
\ref{nondetectFIR}.  Table \ref{detectOH} lists the OH emission
properties and 1.4 GHz flux density of the OH detections in the following
format.
Column (1):  {\it IRAS} FSC name.
Column (2):  Measured heliocentric velocity of the 1667.359 MHz 
line, defined by the center of the FWHM of the line.  The uncertainty in the
velocity of the line center is estimated assuming an uncertainty of $\pm 1$
channel ($\pm 49$ kHz) on each side of the line.  
Column (3):  On-source integration time in minutes.  
Column (4):  Peak flux density of the 1667 MHz OH line in mJy.
Column (5):  Equivalent width-like measure in MHz.  
$W_{1667}$ is the ratio of the integrated 1667 MHz line flux to its 
peak flux.  Ranges are listed for $W_{1667}$ in cases where the identification
of the 1665 MHz line is unclear.
Column (6):  Observed FWHM of the 1667 MHz OH line in MHz.
Column (7):  Rest frame FWHM of the 1667 MHz OH line in km 
s\minusone.
The rest frame width was calculated from the observed width as
$\Delta v_{rest} = c (1+z) (\Delta \nu_{obs} / \nu_\circ)$.
Column (8):  Hyperfine ratio, defined by $R_H = F_{1667}/F_{1665}$, where 
$F_\nu$ is the integrated flux density across the emission line centered on 
$\nu$.  $R_H = 1.8$ in thermodynamic equilibrium, and increases as the degree
of saturation of masing regions increases.  In many cases, the 1665 MHz
OH line is not apparent, or is blended into the 1667 MHz OH line, and a good
measure of $R_H$ becomes difficult without a model for the line profile.  It is
also not clear that the two lines should have similar profiles, particularly if the
lines are aggregates of many emission regions in different saturation states. 
Some spectra allow a lower limit to be placed on $R_H$, indicated by a 
``greater than''
symbol.  Blended or noisy lines have uncertain values of $R_H$, and are indicated 
by a tilde, but in some cases, separation of the two OH lines is impossible
and no value is listed for $R_H$.  
Column (9):  Logarithm of the FIR luminosity, as in Table \ref{detectFIR}.
Column (10):  Predicted OH luminosity, $\log L_{OH}^{pred}$, as in Table 
\ref{nondetectOH}.
Column (11):  Logarithm of the measured isotropic OH line luminosity, 
which includes the 
integrated flux density of both the 1667.359 and the 1665.4018 MHz lines.
Note that $L_{OH}^{pred}$ is generally 
less than the actual \LOH detected (23 out of 24 detections).
Column (12):  1.4 GHz continuum fluxes, from the NVSS.
If no continuum source lies within 30$\arcsec$ of the {\it IRAS} 
coordinates, an upper limit of 5.0 mJy is listed.  

The spectra of the OH detections are presented in Figure \ref{spectra}.  
The abscissae and inset redshifts refer to the optical heliocentric 
velocity, and the arrows indicate the expected velocity of the 1667.359 
({\it left}) and 1665.4018 ({\it right}) MHz lines based on the optical 
redshift, with error bars indicating
the uncertainty in the redshift.  The spectra refer to  1667.359 MHz as 
the rest
frequency for the velocity scale.  Spectra have had the dotted baselines
subtracted, and the baselines have been shifted in 
absolute flux density such that the central channel has value zero.  
The small frame below each spectrum shows a weights spectrum, indicating
the fractional number of records used to form the final spectrum after
the RFI rejection procedure (\S \ref{obs}).  
Channels with weights close to unity are ``good''
channels, whereas channels with lower than average weight are influenced by
time-variable RFI and are thus suspect.  The weights spectra are presented
to indicate confidence in various spectral features, but note that often
the RFI rejection procedure does a good job of cleaning channels and that
channels with $\sim10\%$ rejected records may be completely reliable (this
is, after all, the point of the RFI cleaning procedure).

In order to quantitatively identify dubious 1665 MHz OH line
detections, we compute the autocorrelation function (ACF) of each spectrum and 
locate the secondary peak (the primary peak corresponds to zero offset, or
perfect correlation).   Any correspondence of features between the two main 
OH lines will enhance the second autocorrelation peak and allow us to 
unambiguously identify 1665 MHz lines based not strictly on spectral 
location and peak flux, but on line shape as well.
The secondary peak in the ACF of each spectrum, when present, 
is indicated by a small solid line over the spectra in Figure \ref{spectra}.  
We expect 
the offset of the secondary peak to be equal to the separation
of the two main OH lines, properly redshifted:  (1.9572 MHz)$/(1+z)$.  The 
{\it expected} location of the secondary ACF peak is indicated
in Figure \ref{spectra} by a small dashed line over each spectrum.  
Both the expected and actual secondary peak positions are plotted 
offset with respect to the center of the 1667 MHz line, as defined by the 
center of the FWHM, rather than the peak flux.  

We examined the Digitized Sky Survey\footnote{Based on 
photographic data obtained using Oschin Schmidt Telescope on Palomar 
Mountain. The Palomar Observatory Sky Survey was funded by the National 
Geographic Society. The Oschin Schmidt Telescope is operated by the 
California Institute of Technology and Palomar Observatory. The plates 
were processed into the present compressed digital format with their
permission. The Digitized Sky Survey was produced at the Space Telescope 
Science Institute (STScI) under U.S. Government grant NAG W-2166.}
(DSS) images of each new OH detection.   The OHM hosts 
are generally faint, unresolved, and unremarkable in the DSS unless otherwise
noted in the discussion of individual sources below.  We also performed
an exhaustive literature search for each new OHM, and searched the Hubble
Space Telescope (HST) archives for fields containing OHM hosts.  All 
relevant data are included in the discussions below.
The weights spectra
are generally clean across the OH line profiles, unless specifically noted.
We make some observations and measurements specific to individual OH detections
as follows.

\noindent{\bf 04121+0223: }
The optical redshift and observed line velocities are in good agreement for
this source.  The predicted location of the 1665 MHz line from the 1667 MHz 
centroid and from the strong second peak in the spectral ACF are in good agreement
and correspond to a significant spectral feature (a $2\sigma$ feature in 
single-channel flux density, but it is broad, with an integrated flux 
density of of 0.61 mJy MHz and FWHM of 1.54 MHz).  
There are clearly two distinct emission features 
in the 1667 MHz line:  one fairly broad ($135 \pm 12$ km s\minusone\ rest-frame
FWHM) and one narrow ($37 \pm 12$ km s\minusone).  
The DSS image of {\it IRAS} 04121+0223 shows a slightly extended host galaxy.

\noindent{\bf 07163+0817: }
The 1667 MHz line of this source shows
an extremely narrow emission spike which appears to be spectrally unresolved 
($24 \pm 12$ km s\minusone\ rest-frame FWHM).  The two 1667 MHz emission peaks 
may correspond to a toroidal emission configuration or to two masing nuclei in 
the merger.  
We use the centroid of the main 1667 MHz line complex (excluding the narrow
emission spike) as the reference for the
ACF and the predicted offset of the 1665 MHz line.  The ACF is dominated by 
peaks produced when the strong OH emission spike corresponds with any other 
narrow spike in the spectrum.  The predicted 1665 MHz line
velocity seems to correspond roughly with a $3 \sigma$ spectral feature.  Assuming
that this feature is the 1665 MHz line, we obtain $R_H \sim 5.5$.  
The identification
of the 1665 MHz line is uncertain, which is indicated by the tilde in Table 
\ref{detectFIR}.  
The weights spectrum indicates some narrow RFI at 33500 km s\minusone\ which
may mildly influence the integrated 1667 MHz line flux.

\noindent{\bf 07572+0533: }
The ACF and offset predictions for the 1665 MHz line velocity show good 
agreement, 
but there is no significant corresponding spectral feature in this OH 
spectrum.  We compute a lower
limit on the hyperfine ratio assuming a square profile of width equal to 
the 1667
MHz line width and height $1\sigma$ to obtain $R_H \geq 10.4$.  There 
appears to
be a second OH line (a $3\sigma$ detection) offset blueward by 400 km 
s\minusone\ 
in the rest frame from the center of the
main 1667 MHz line.  This line is included in the computation of the isotropic
OH line luminosity for this source.  The features at 57300, 58150, and 58600
km s\minusone\ are identified with RFI.  The weights array is otherwise
clean across the OH spectrum.

\begin{figure}[ht]
\plotone{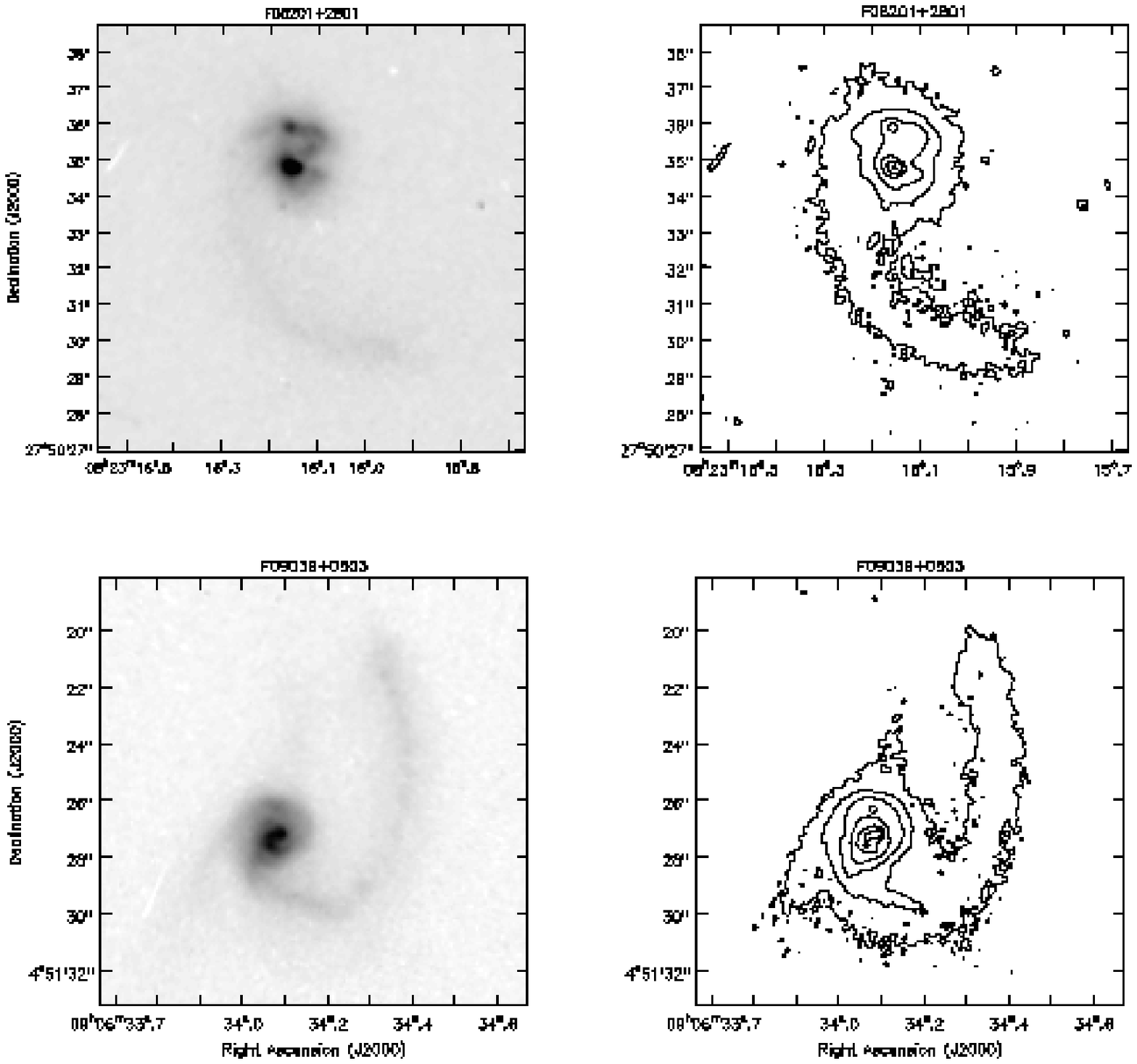}
\caption{HST WFPC2 archive images of {\it IRAS} 08201+2801 
and {\it IRAS} 09039+0503
(800 s exposures; F814W filter; $\lambda = 7940$\AA, $\Delta\lambda = 
1531$\AA) reveal morphological signatures of interacting galaxies.
The greyscale is linear, and contours are incremented by factors of 2.
{\it IRAS} 08021+2801 shows two nuclei and an extended tidal tail.
{\it IRAS} 09039+0503 has a double nucleus and three tidal tails.
\label{F08201img}}
\end{figure}

\noindent{\bf 08201+2801: }
The nucleus of this ULIRG was classified by \citet{kim98b} as a starburst.
The spectrum around 52200 km s\minusone\ has been masked to remove incompletely
subtracted strong Galactic \ion{H}{1} emission.  The ACF shows no second peak, 
but the predicted velocity of the 1665 MHz line corresponds to a feature of
similar width to the main 1667 MHz line.  There is a similar shoulder, however,
on the blue side of the main line, which may indicate that the feature we 
identify as the 1665 MHz line is instead part of a high-velocity complex of 
low-level 1667 MHz emission.  If we assume that the feature on the red side
is the 1665 MHz line, then we obtain a lower bound on the hyperfine 
ratio:  $R_H \geq 8.2$.  Note that there is a mild depression in the weights
spectrum near the expected 1665 MHz line, but we are satisfied that the 
RFI was successfully removed. The spectrum shows an absorption feature redward 
of the OH emission complex (the 
NVSS 1.4 GHz continuum flux of this source is 16.7 mJy; see Condon et al. 1998)
which may indicate infall of molecular gas.  This
would be consistent with the picture of ULIRGs as merging systems in which
gas and dust becomes concentrated into the merging nuclei by tidal angular 
momentum loss and subsequent infall.  If the line is absorption in 1667 MHz, 
then the absorption minimum has a rest frame offset from the 1667 MHz 
emission line center of 
750 km s\minusone. If the line is absorption in 1665 MHz, the offset 
from the 1665 MHz emission line is 380 km s\minusone.  Alternatively, 
the absorption could occur in a physically distinct region from the 
masing region, along a different line of sight, such as in another nucleus.
A WFPC2 HST archive image\footnote{Based on observations 
made with the NASA/ESA Hubble Space Telescope, obtained from the data 
archive at the Space Telescope Science Institute. STScI is operated by 
the Association of Universities for Research in Astronomy, Inc. under 
NASA contract NAS 5-26555.}
(F814W, 800 s) shows interesting morphology, with two
nuclei connected by arcs, and a single kinked tidal tail (see Fig.\ 
\ref{F08201img}).  
This is clearly an advanced merger, and we can measure a separation
between the two nuclei for an indication of merger stage.
The nuclear separation is 1.09\arcsec, corresponding to a projected 
distance of 2.83 kpc ($R = (1+z)^{-2} D_L \Delta\theta$).  
If the absorption feature at 51075 km s\minusone\ is associated
with one of the two nuclei, while the emission is associated with the 
other nucleus, then we estimate a rough enclosed mass of 
$M(r) \simeq v^2R/G = 3.7\times 10^{11}$ M$_\odot$ 
and a naive crossing time of $t_{cross} \simeq R/v = 3.7\times 10^6$ years.

\noindent{\bf 08279+0956: } 
There is good agreement in this source between the OH line velocities, the optical 
redshift predictions for the two OH lines, the predicted location of the 1665 MHz
line from the ACF, and the predicted location of the 1665 MHz line from the 
1667 MHz line center.  The isotropic OH line luminosity of this source is nearly
an order of magnitude larger than we predict from \LFIR.  The weights spectrum is
clean across the OH lines.  The ``absorption'' feature at 61200 km s\minusone\
is produced by a resonance in the wide L-band feed at Arecibo.  

\noindent{\bf 08449+2332: } 
No significant spectral feature corresponds to the predicted velocity of the
1665 MHz line from the 1667 MHz line centroid or from the ACF second peak.  
The emission profile shows a narrow line atop a broader base of emission, 
spanning 760 km s\minusone\ at $10\%$ of the peak flux density in the rest 
frame.  Although the
isotropic OH luminosity is somewhat low, it is still higher than the value 
predicted from $L_{FIR}$.  A lower bound is calculated for the hyperfine ratio:
$R_H \geq F_{1667}/(1\sigma \Delta\nu_{1667}) = 11.0$.  The weights array is 
fairly clean across the OH line complex.

\noindent{\bf 08474+1813: }
This source shows a broad 1667 MHz emission line which may be blended
with the 1665 MHz line.  The ACF has no secondary peak, but the optical 
redshift and the center of the 1667 MHz line both indicate a 1665 MHz line
velocity corresponding to the blended feature redward of the bulk of the 
emission.  Using the assumption that this is the 1665 MHz line as a lower 
limit, $R_H \geq 3.0$.  The equivalent width-like measure, $W_{1667}$, 
is assigned a range of values which bracket the limiting cases for the
presence of the 1665 MHz line.  The weights spectrum shows a mild dip near the 
possible 1665 MHz line, indicating that the 1665 MHz line identification
and measurements should be interpreted with caution.

\noindent{\bf 09039+0503: } 
The nucleus of this OHM host is classified by \citet{vei99} as a LINER.  
The ACF and 1667 MHz line center predictions are in excellent agreement 
with a spectral 
feature we identify as the 1665 MHz line.  Note that there is another emission
feature blueward of the main 1667 MHz line which may be associated with another
nucleus or with an outflow or molecular gas torus.  All measurements of the 
1667 MHz line
include this emission.  The weights array is clean, except for a small 
depression in the 1665 MHz line velocity. 
HST WFPC2 archive images (F814W, 800 s) of this source reveal a remarkable
morphology including two nuclei and three tidal tails (see Fig.\ 
\ref{F08201img}).  
The nuclear separation is 0.56\arcsec, corresponding to a projected 
distance of 1.15 kpc.  If the blue
OH peak at 37300 km s\minusone\ 
is associated with a different nucleus from the main peak, then 
the velocity difference is 465 km s\minusone. 
This gives rough estimates for a crossing time of $2.4\times10^6$ 
years and an enclosed mass of $5.7\times10^{10} M_\odot$.

\noindent{\bf 09531+1430: }  
The OH spectrum of this object shows two peaks in 1667 MHz emission with
a rest frame velocity separation of 410 km s\minusone.  The 
redder of the two peaks coincides with the optical redshift.  
One interpretation of this OH spectrum could be that the two
lines are emitted from two nuclei in a merger, one of which is optically 
dominant or unobscured.  The OH emission is integrated
across the entire line complex to obtain $L_{OH}$, but only 
the 1667 and 1665 MHz
peaks on the red side are used to compute $W_{1667}$ and $R_H$.
The ACF and the centroid of the red 1667 MHz line agree on the predicted 
location
of the 1665 MHz line, and the prediction coincides with a conspicuous spectral
feature.  There is a significant depression in the weights
spectrum at the location of a potential 1665 MHz line --- 
the hyperfine ratio is thus suspect.  
It is likely that the red 1667 MHz line contains the 1665 MHz emission 
associated with
the bluest emission line, and is hence artificially strong depending on the
hyperfine ratio of the blue complex.
Assuming that this is the 1665 MHz line associated with the 
strongest 
emission line, but keeping in mind the likely boosting of the 1667 MHz 
flux by overlap with the blue 1665 MHz line, we obtain an uncertain 
$R_H\sim 3.4$.  
The FWHM is measured from the strongest OH line in the spectrum (the red
1667 MHz line).
The rest frame width of the entire OH emission complex at 10$\%$ peak 
flux density is 1040 km s\minusone.

\noindent{\bf 09539+0857: } 
The host of this OHM is classified by \citet{vei99} as a LINER.
The ACF is extremely smooth and has no second peak due to the highly 
blended OH lines, but the predictions of the optical redshift and
the 1667 MHz line center for the 1665 MHz line velocity are congruent with 
a very strong feature we identify as the 1665 MHz line.  The hyperfine ratio
is unusually small for the megamaser sample:  $R_H = 2.5$.  Note that \LOH is
more than an order of magnitude larger than $L_{OH}^{pred}$.  The weights
spectrum is very smooth across the OH spectrum.

\noindent{\bf 10339+1548: } 
The baseline of this OH spectrum shows mild standing waves that frustrate 
detection of the 1665 MHz OH line.  There is no second peak in the ACF.  
The position of the single 1667 MHz spike indicates a 
broad feature to be the 1665 MHz line, but this feature cannot be distinguished
from a standing wave in the bandpass.  We use this feature, however, to set a
lower limit on the hyperfine ratio:  $R_H \gtrsim 14.5$.  This OHM shows a single 
narrow peak which is probably spectrally unresolved.  
The feature at 61200 km s\minusone\ is associated with RFI, as is the sharp
dip on the blue side of the OH line at 58800 km s\minusone.
The weights spectrum is otherwise fairly clean across the bandpass.

\noindent{\bf 11028+3130: } 
The nucleus of this OHM host is classified by \citet{kim98b} as a LINER. 
A marginally significant
spectral feature has a flux centroid matching the predicted location 
of the 1665 MHz line from the ACF and the 1667 MHz line centroid (the two
predictions overlap on the plot).  We use this feature to compute the hyperfine
ratio, which has a typical OHM value of $R_H = 5.5$.  
The optical redshift is in excellent agreement with the OH redshift and the 
feature identified as the 1665 MHz line.  Weights spectrum is fairly clean 
across the OH spectrum.

\noindent{\bf 11524+1058: } 
The spectrum near 52200 km s\minusone\ has been masked to remove incompletely
subtracted strong Galactic \ion{H}{1} emission.  The ACF shows no second peak, 
but the predicted location of the 1665 MHz line from the 1667 MHz line centroid 
identifies a significant spectral feature as the 1665 MHz line which we use
to compute $R_H\simeq4.9$.  The optical 
redshift is consistent with the OH redshift.  The 1667 MHz line is double-peaked, 
and the peaks have nearly equal flux density and a rest frame separation of 130
km s\minusone.  The weights array is smooth across the OH spectrum.

\noindent{\bf 12032+1707: }  
This is an extremely complicated and luminous OH megamaser (it is in fact a 
``gigamaser'').  {\it IRAS} F12032+1707 is among the most luminous 
OHMs detected, comparable to {\it IRAS} 14070+0525 \citep{baa92b} and
{\it IRAS} 20100-4156 (Staveley-Smith et al. 1992).
The width of the line complex at $10\%$ peak flux is 1470 km 
s\minusone.  There appears to be a long high-velocity tail to the OH emission.
There is undoubtedly blending of the 1667 and 1665 MHz lines in this spectrum, 
and disentangling the two is impossible without more information.  The centroid
of the emission profile and the optical redshift predict a 1665 MHz line which
would be blended with the observed 1667 MHz line emission.  There is no second 
peak in the ACF.  We make no estimate of the hyperfine ratio, and $W_{1667}$
is computed from the entire line profile.  
The computed FWHM uses the highest flux density 
spectral feature as the maximum, and includes the entire 1667 MHz complex.  
The optical redshift corresponds with the centroid of the broad emission complex.
It is possible that the two line complexes emitted from this
source (the narrow strong single line, and the broad flat-topped complex) originate
in two nuclei with rest frame separation of 570 km s\minusone\ 
(centroid-to-centroid).  This OHM host also has the largest 1.4 GHz continuum 
flux density (28.7 mJy) of any of the new OHMs in the survey to date.  
This remarkable object was classified by \citet{vei99} as a LINER.

\noindent{\bf 15224+1033: } 
This OHM has an extremely sharp central peak 
and broad red and blue wings on the main line.  There are 
also two significant minor peaks in the emission profile.  
The ACF has a steplike structure,
but shows no strong secondary peak.  The 1667 MHz line centroid and the 
optical redshift make similar predictions for the 
1665 MHz line velocity, but no significant feature is present
in the spectrum.  The emission spectrum at the expected 1665 MHz line velocity 
is part of the broad red wing in the emission profile, with a similar shape
to the blue wing.  If we assume that the red wing is in fact 1665 MHz emission, 
then we obtain a lower bound on the hyperfine ratio:  $R_H \geq 9.5$.  
$W_{1667}$ is
given a range of values, under the two limits of maximum and no 1665 MHz emission.
The OH luminosity of this source is more than an order of magnitude greater than 
$L_{OH}^{pred}$.  

\noindent{\bf 15587+1609: } 
Identification of the 1665 MHz line is unambiguous in this OHM.  The ACF and the 
1667 MHz line centroid predictions for the 1665 MHz line agree and correspond
to a highly significant spectral line.  There is some offset between
the optical redshift and the OH redshift; the 1667 MHz line centroid is bluer
than the optical redshift by 300 km s\minusone, a $1.5\sigma$ departure from 
the optical redshift.  The OH luminosity is more than an order of magnitude
greater than $L_{OH}^{pred}$.  

\noindent{\bf 16100+2527: } 
The ACF and 1667 MHz line centroid predictions for the 1665 MHz line are
in excellent agreement (the two predictions overlap in the plot), and 
correspond to a significant spectral line feature.  The optical redshift
is significantly different from the OH redshift:  770 km s\minusone.  This
is a $3\sigma$ departure from the optical redshift.  We conclude that either
the optical redshift is erroneous or the optically bright portion of this 
merger is kinematically distinct from the masing region.  The former
hypothesis is supported by Palomar 5m spectra which provide redshift 
determinations from many optical lines in agreement with the OH redshift.
These data will be published in a subsequent paper.
This source shows an unusually lopsided OH line profile in both emission
lines.  The blue side of each line shows a slow drop-off while the drop-off
on the red side is quite abrupt from the peak to zero flux density.  

\noindent{\bf 16255+2801: } 
The {\it IRAS} faint source associated with this OHM has been misidentified
by \citet{con98b} as a planetary nebula based on an NVSS survey detection
\citep{con98}.  In fact, the $95\%$ confidence position ellipses from the
NVSS and the {\it IRAS} FSC do not overlap.  We obtained optical spectra
at the Palomar 5m telescope of several sources in the field of 
{\it IRAS} 16255+2801 and found only one source with a redshift in agreement 
with the OH and PSCz redshifts.  The Palomar observations will
be discussed in a subsequent paper.  The true J2000 coordinates of this source
are $16^h 27^m 38.1^s$ $+27^\circ 54^m 52^s$, which exactly
correspond with a FIRST point source (1.1 mJy; White et al. 1997).  
The optical and OH redshifts are somewhat different, although within the 
uncertainty of the optical redshift.  
The small feature blueward of the main 1667 MHz peak is included in
the total OH flux measure.  Although there is no credible second peak in the ACF
corresponding to the 1665 MHz line, there is a $3.5\sigma$ feature at 
roughly the velocity predicted by the 1667 MHz line centroid.  This line
is not significantly different from other large deviations in the bandpass, 
and does not indicate an unambiguous detection of the 1665 MHz line.  
This feature provides a lower bound on the hyperfine ratio:  $R_H \geq 13.7$.  

\noindent{\bf 22055+3024: } 
The spectrum of this OHM is contaminated by a digital radio satellite signal 
across the bandpass, notably conspicuous at 36250, 37000, 38150, 40050, and
40250 km s\minusone, although the RFI is quite narrow, mostly constant 
in time, and does not significantly affect the OH spectrum.
One polarization of a 4-minute integration was omitted from the 
weighted average
due to strong RFI (not associated with the satellite RFI).  
The weights spectrum is fairly clean across the OH spectrum, 
although there is some time variability in a narrow band between 
the main line and the small blue peak, 
causing rejection of $10\%$ of the records.  
There are significant peaks redward and blueward of the main 1667 MHz line.
We identify the red peak as the 1665 MHz line, which shows good correspondence
between the spectral features, the ACF secondary peak, the 1667 MHz line centroid
prediction, and the optical redshift.  The blue peak is probably 1667 MHz 
emission, 
although it could be affected somewhat by the mild standing waves evident in the
bandpass.  The blue peak is included in measurements of the
1667 MHz integrated flux.

\noindent{\bf 23019+3405: } 
The DSS image of this OHM host shows a galaxy pair, although 
the {\it IRAS} FSC uncertainty ellipse
does not unambiguously identify the pair as the source of FIR flux.  
This OHM spectrum consists of a single narrow emission line.  There are no
highly significant secondary peaks, wings or shoulders in the OH spectrum.  
The ACF and the 1667 MHz line predict a 1665 MHz line velocity in agreement
with the optical redshift and a spectral feature, but the feature is not
significant (a $2.9\sigma$ peak flux density feature).  We use this feature
to obtain a lower bound on the hyperfine ratio:  $R_H \geq 15.6$.  
The weights spectrum is 
flat across the OH spectrum, but shows significant RFI features above 32800 km
s\minusone. 

\noindent{\bf 23028+0725: } 
The ACF does not show a secondary peak --- only ``broad shoulders'' due
to wide, blended OH lines --- although the 1665 MHz line is obvious
in this source.  The 1665 MHz line corresponds to the predicted offset from the
1667 MHz line centroid.  The OH and optical redshifts differ somewhat, but 
agree to within the uncertainty in the optical redshift.  The 1665 MHz line
is quite strong in this source, and the hyperfine ratio almost reaches the
thermodynamic equilibrium value of 1.8:  $R_H = 1.9$.  
The weights spectrum reveals narrow-band RFI (spanning 0.3 MHz) at 
45000 km s\minusone\ which causes a rejection of $\sim4\%$ of the records.  
This RFI does not significantly affect the integrated flux of the
1665 MHz line.

\noindent{\bf 23129+2548: } 
The spectrum of this OHM has been masked around 52000 km s\minusone\ 
to remove incompletely
subtracted strong Galactic \ion{H}{1} emission.  The 1667 and 1665 MHz lines 
are blended in this source, and the ACF has no secondary peak.  The 
1665 MHz line velocity predictions from the 1667 MHz line centroid and the
optical redshift
are in good agreement, but there is no way to disentangle the two emission lines
in this OH spectrum.  $W_{1667}$ is computed from the entire line flux.  
The nucleus of this OHM host is classified by \citet{vei99} as a LINER.

\noindent{\bf 23199+0123: } 
The ACF and the 1667 MHz line centroid identify a marginally significant
spectral feature as the 1665 MHz line.  The optical redshift differs from
the OH redshift by $1.2\sigma$.  The bandpass for this source has 
strong curvature, as indicated by the dotted baseline.
Although this megamaser is fairly weak compared
to the rest of the sample, it has an isotropic OH line luminosity 
significantly greater than $L_{OH}^{pred}$.  

\noindent{\bf 23234+0946: }
The host of this OHM is classified by \citet{vei99} as a LINER.  
The DSS image shows a resolved, non-axisymmetric galaxy with a bright 
nucleus surrounded by extended emission.
This source could use a longer integration time for a higher signal-to-noise
spectrum.  A prediction of the 1665 MHz line velocity is strongly subject to 
an uncertain FWHM of the 1667 MHz line.  The peak redward of the main 1667 MHz
line is probably the 1665 MHz line and is used to compute the
hyperfine ratio.  The OH and optical redshifts differ by 120 km s\minusone, which
is a $4.8\sigma$ discrepancy in the optical redshift.  
It is likely that the uncertainty in the optical
redshift is a significant underestimate.  There is a significant peak blueward
of the main peak, which may be 1667 MHz emission.  This blue peak is included
in the 1667 MHz line flux measure.  The weights spectrum shows a loss 
of $\sim 2.5\%$ of the
records due to RFI rejection in a flat band extending across the main 1667 and
1665 MHz lines, but excluding the blue peak.  Examination of individual records
reveals sporadic weak boxcar-shaped RFI.  We are satisfied that the RFI rejection 
procedure has removed this feature from the final OH spectrum.

\section{Discussion}\label{discussion}

Analysis will 
be reserved for a completed survey, but several trends are already
evident:  (1) there is a strong positive correlation between the OHM 
fraction in luminous IR galaxies and the FIR luminosity;  (2)
there is a positive correlation between the isotropic OH line luminosity
and $L_{FIR}$, but the relationship shows strong scatter, probably
related to viewing geometry and maser saturation states.  Both of these 
trends have been observed in lower redshift samples by \citet{baa89}, 
\citet{sta92}, and others.  
There are two additional trends seen in the sample which are currently 
under investigation with optical telescopes:
(1) the majority of the OHM hosts with optical spectral classifications
available in the literature are LINERs; (2) all of the OHM hosts with 
available HST archive images show multiple nuclei with small physical 
separations.  

We require a detailed understanding of the relationship between merging 
galaxies and the OH megamaser phenomenon in order to obtain a galaxy merger 
rate from OHM surveys.  Hence, one needs to 
understand the observed trends in the survey sample.  Analysis of the
completed survey will include a detailed description of the survey biases
and completeness, and a re-evaluation of the $L_{OH}$-\LFIR relationship, 
taking into account the survey biases and the confidence in nondetections.
Follow-up work related to the survey is underway, including optical 
spectroscopic identification of OHM hosts, variability studies, and high
angular resolution imaging.  We are also conducting a survey of AGN for
OHMs in order to confirm the exclusive connection between OH megamasers
and major galaxy mergers.  Finally, we plan to conduct deep surveys
for OH at
high redshifts and measure the merger rate of galaxies as a function of
cosmic time.




\acknowledgements
 
The authors are very grateful to Will Saunders for access to the PSCz catalog
and to the excellent staff of NAIC for observing assistance and support.  
This research was supported by Space Science Institute archival grant 
8373 and made use of the NASA/IPAC Extragalactic Database (NED) 
which is operated by the Jet Propulsion Laboratory, California
Institute of Technology, under contract with the National Aeronautics 
and Space Administration.  
We acknowledge the use of NASA's SkyView facility 
(http://skyview.gsfc.nasa.gov) located at NASA Goddard Space Flight Center.

\clearpage

\begin{figure}[ht]
\plotone{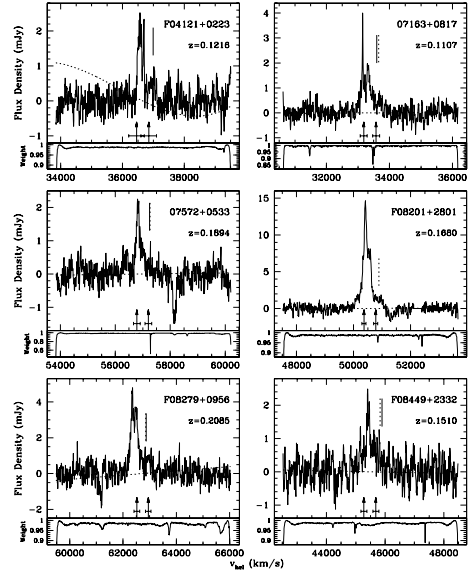}
\caption{New OH megamasers discovered in (U)LIRGs.  
Abscissae and inset redshifts refer to the optical heliocentric velocity.
Spectra use the 1667.359 MHz line as the rest frequency for the 
velocity scale.  
Arrows indicate the expected velocity of the 1667.359 (left)
and 1665.4018 (right) MHz lines based on the optical redshift, with 
error bars indicating the uncertainty in the redshift.  Solid vertical
lines indicate the location of the secondary maximum in the autocorrelation
function, and dashed vertical lines indicate the expected position of the 
1665 MHz line, based on the centroid of the 1667 MHz line; a match between
the two indicates a possible detection of the 1665 MHz line.  The 
dotted baselines indicate the shape (but not the absolute magnitude) 
of the baselines subtracted from the calibrated spectra.
The small frame below each spectrum shows the ``weights'' spectrum, indicating
the fractional number of RFI-free records averaged in each channel.
The properties of these megamasers are listed in Tables \ref{detectFIR}
and \ref{detectOH}.\label{spectra}}
\end{figure}

\begin{figure}
\plotone{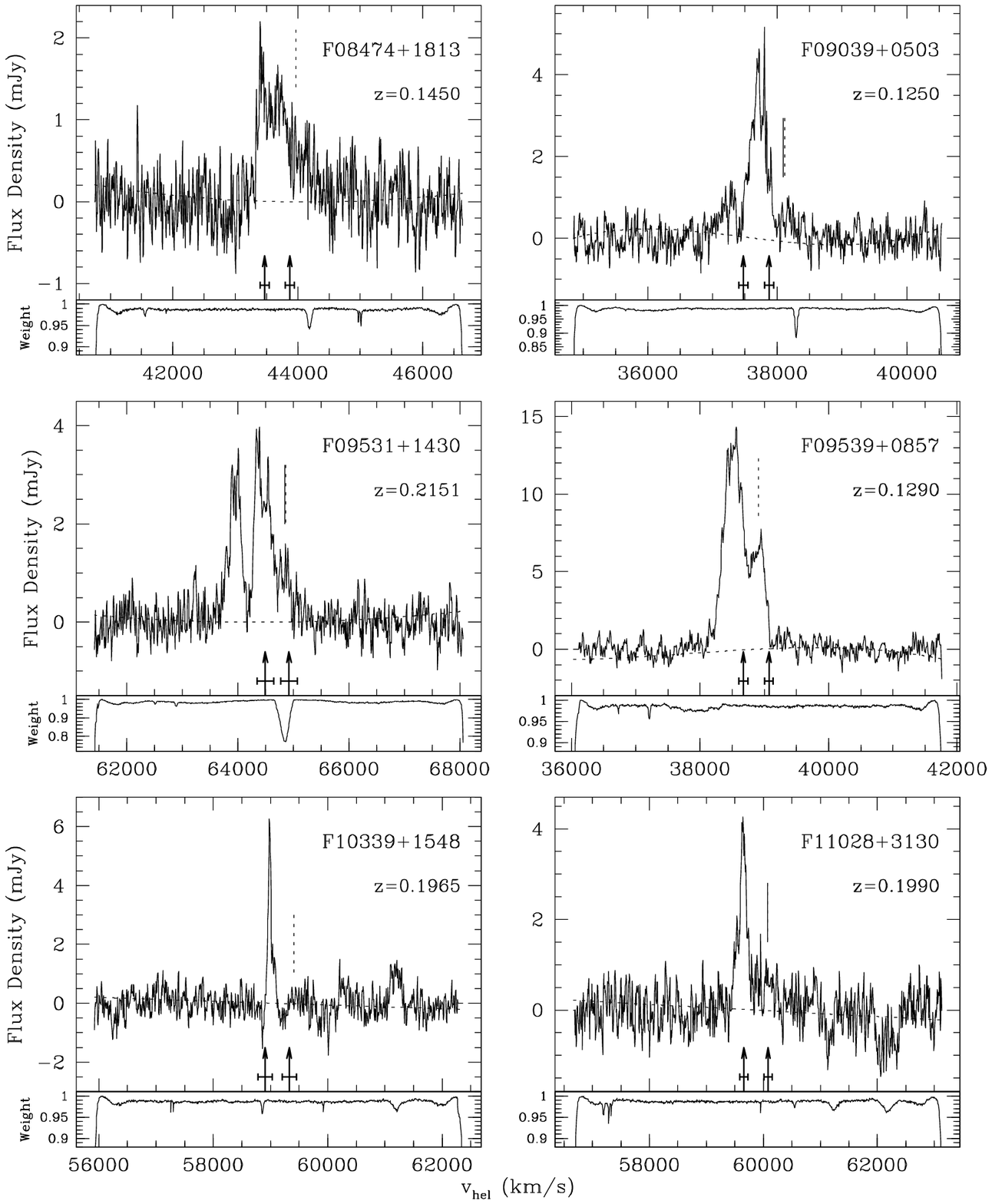}
\figurenum{2}
\caption{{\it continued.}}
\end{figure}

\begin{figure}
\plotone{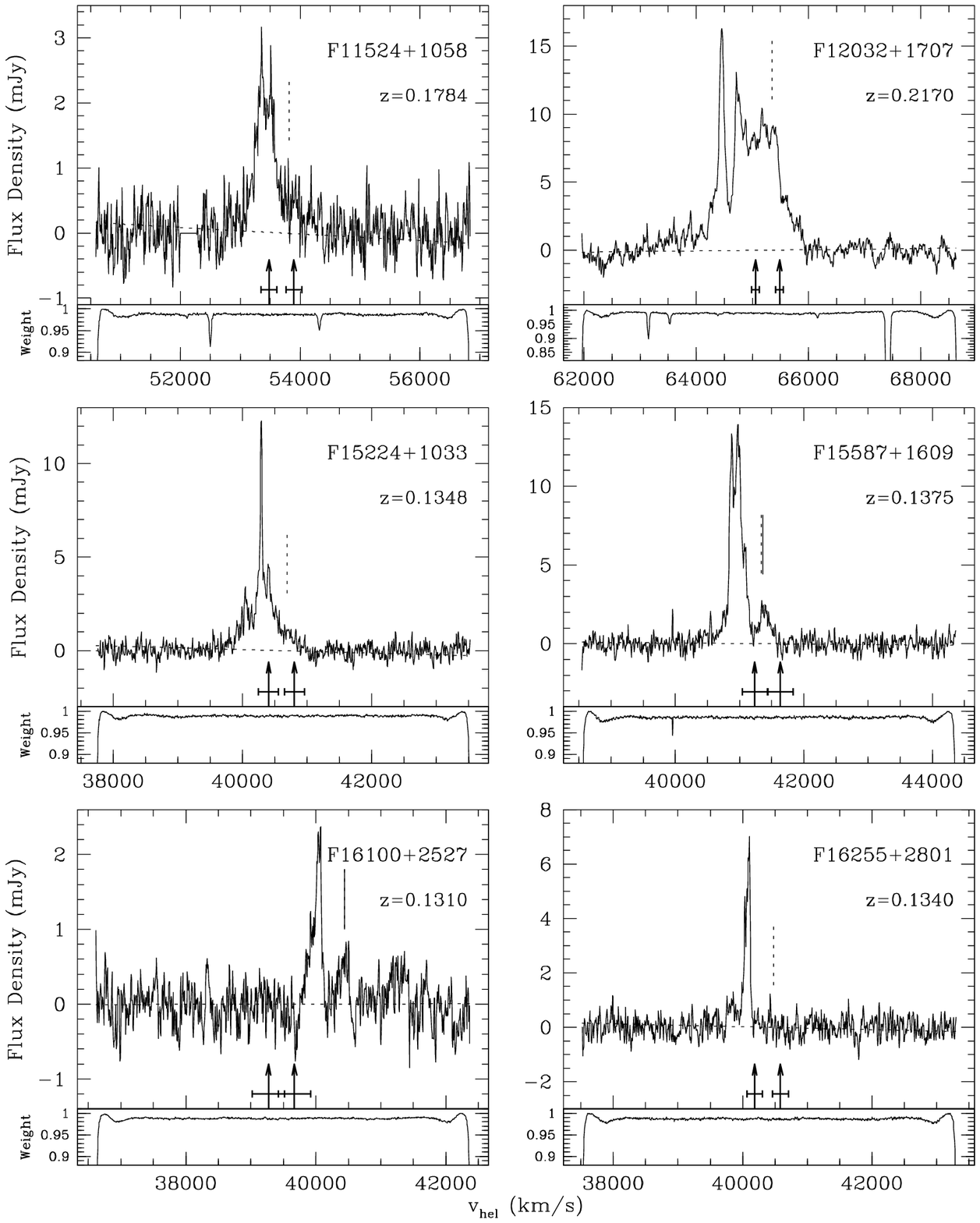}
\figurenum{2}
\caption{{\it continued.}}
\end{figure}

\begin{figure}
\plotone{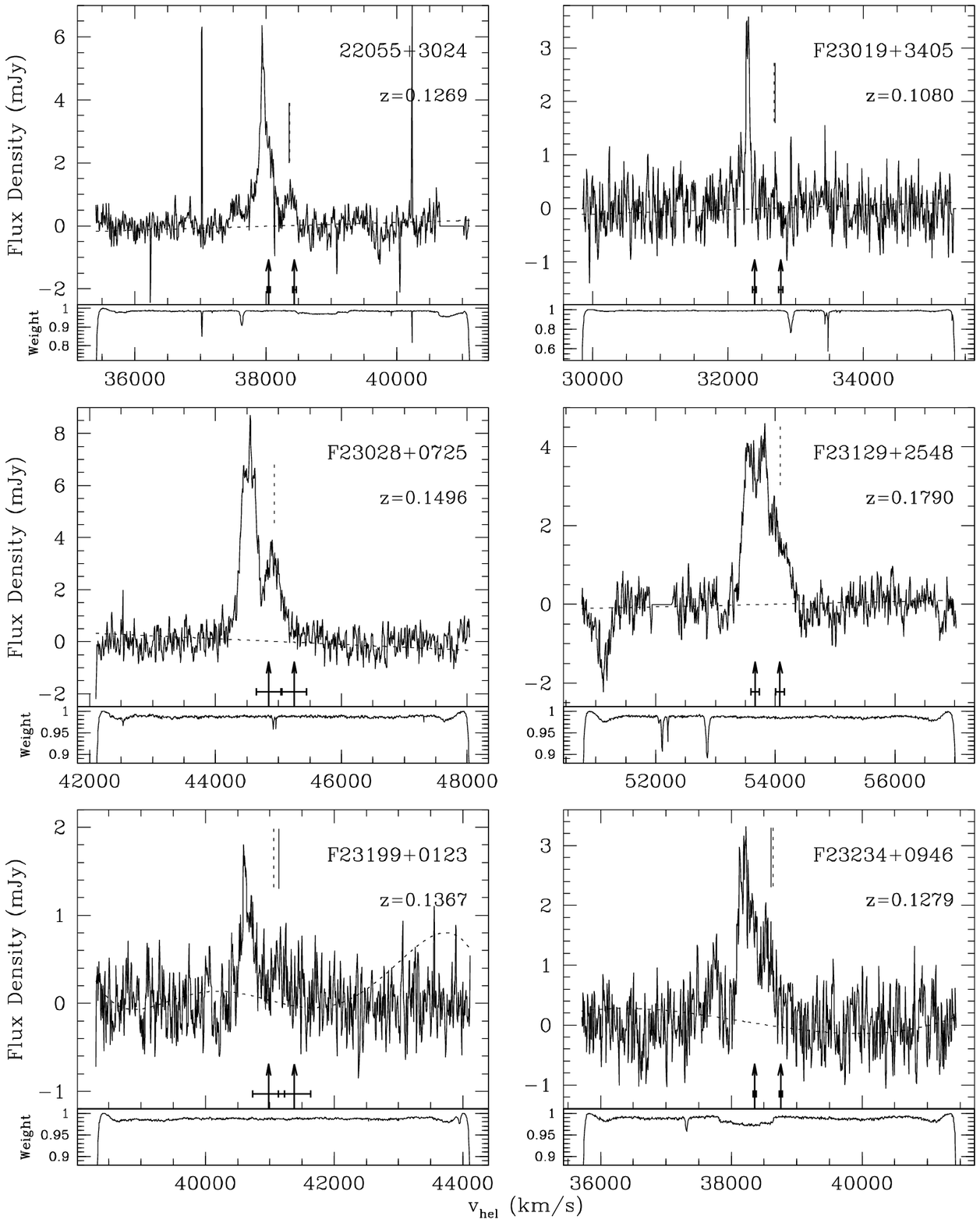}
\figurenum{2}
\caption{{\it continued.}}
\end{figure}

\clearpage

\begin{deluxetable}{cccccrrcrcc} 
\tabletypesize{\scriptsize}
\tablecaption{OH Non-Detections:  Optical Redshifts and FIR Properties\label{nondetectFIR}}
\tablewidth{0pt}
\tablehead{
\colhead{{\it IRAS} Name} &  \colhead{$\alpha$} & \colhead{$\delta$} & \colhead{$z_\odot$}  
& \colhead{Ref}& \colhead{$v_\odot$} & \colhead{$v_{CMB}$} 
& \colhead{$D_L$} & \colhead{$f_{60}$} & \colhead{$f_{100}$} & 
\colhead{$\log L_{FIR}$} \\
\colhead{FSC} & \colhead{B1950} & \colhead{B1950} &  & \colhead{} & \colhead{km/s} &
\colhead{km/s} &  \colhead{$h^{-1}_{75}$ Mpc} &  \colhead{Jy} &  
\colhead{Jy} & \colhead{$h^{-2}_{75} L_\odot$} \\
\colhead{(1)}& \colhead{(2)}& \colhead{(3)}& \colhead{(4)}& \colhead{(5)}& 
\colhead{(6)}& \colhead{(7)}& \colhead{(8)}& \colhead{(9)}& \colhead{(10)}& 
\colhead{(11)}
 }
\startdata
00051+2657 & 00 05 06.8 & +26 57 08 & 0.1254 & 1 & 37587(145)
& 37250(146) & 528(2) & 0.649(\phn65) & 1.75(21) & 11.58 \\
03248+1756 & 03 24 51.6 & +17 57 00 & 0.1257 & 1 & 37698(213)
& 37531(216) & 532(3) & 0.677(\phn68) & 1.24(27) & 11.52 \\
03250+1606 & 03 25 00.2 & +16 06 34 & 0.1290 & 2 & 38673(\phn70)
& 38506(\phn78) & 546(1) & 1.381(\phn83) & 1.77(30) & 11.80 \\
03477+2611 & 03 47 43.3 & +26 11 55 & 0.1494 & 1 & 44779(196)
& 44645(199) & 640(3) & 0.711(\phn50) & 1.36(23) & 11.71 \\
04229+0056 & 04 22 55.7 & +00 56 19 & 0.1530 & 13 & 45868(150)
& 45794(154) & 657(2) & 0.494(\phn44) & $<1.70$ & 11.34--11.71 \\
04479+0616 & 04 47 55.1 & +06 16 32 & 0.1158 & 1 & 34715(119)
& 34676(125) & 489(2) & 1.085(119) & 1.93(23) & 11.65 \\
06268+3509 & 06 26 52.3 & +35 09 57 & 0.1698 & 3 & 50904(250)
& 50978(253) & 737(4) & 0.936(\phn75) & 1.09(14) & 11.88 \\
06368+2812 & 06 36 48.5 & +28 12 39 & 0.1249 & 3 & 37444(250)
& 37543(253) & 532(4) & 1.190(107) & 2.01(20) & 11.76 \\
06561+1902 & 06 56 10.5 & +19 02 26 & 0.1882 & 3 & 56420(250)
& 56561(253) & 825(4) & 1.010(\phn81) & 1.32(12) & 12.02 \\
07178+1952 & 07 17 49.1 & +19 52 25 & 0.1148 & 1 & 34411(\phn61)
& 34579(\phn70) & 488(1) & 0.619(\phn50) & 1.62(23) & 11.48 \\
07188+0407 & 07 18 49.2 & +04 07 23 & 0.1035 & 1 & 31036(125)
& 31228(129) & 438(2) & 0.625(\phn56) & $<1.51$ & 11.09--11.37 \\
07241+3052 & 07 24 08.4 & +30 51 58 & 0.1112 & 1 & 33342(122)
& 33493(127) & 472(2) & 0.672(\phn81) & 0.93(22) & 11.37 \\
07328+0457 & 07 32 50.1 & +04 57 05 & 0.1300 & 1 & 38987(110)
& 39197(114) & 557(2) & 1.071(\phn96) & 1.06(14) & 11.67 \\
07381+3215 & 07 38 10.7 & +32 15 11 & 0.1703 & 3 & 51054(250)
& 51217(252) & 741(4) & 0.671(\phn74) & 0.83(18) & 11.75 \\
08003+0734 & 08 00 22.0 & +07 34 01 & 0.1179 & 1 & 35353(154)
& 35593(157) & 503(2) & 0.680(\phn54) & 1.59(25) & 11.52 \\
08007+0711 & 08 00 42.1 & +07 11 57 & 0.1405 & 1 & 42120(105)
& 42361(109) & 605(2) & 0.859(\phn69) & 1.18(13) & 11.69 \\
08012+0125 & 08 01 15.1 & +01 25 18 & 0.2203 & 1 & 66038(118)
& 66286(121) & 982(2) & 0.659(\phn92) & 0.87(10) & 11.99 \\
08122+0505 & 08 12 13.2 & +05 05 30 & 0.1030 & 3 & 30878(300)
& 31135(301) & 437(4) & 1.196(\phn84) & 1.63(16) & 11.55 \\
08132+1628 & 08 13 12.3 & +16 29 02 & 0.1004 & 1 & 30107(114)
& 30347(118) & 425(2) & 0.811(\phn49) & 1.64(16) & 11.43 \\
08147+3137 & 08 14 45.7 & +31 37 50 & 0.1239 & 1 & 37157(\phn65)
& 37359(\phn73) & 529(1) & 0.646(\phn45) & $<0.55$ & 11.27--11.39 \\
08200+1931 & 08 20 01.9 & +19 31 09 & 0.1694 & 1 & 50796(108)
& 51036(112) & 738(2) & 0.866(\phn69) & 0.96(17) & 11.84 \\
08224+1329 & 08 22 27.1 & +13 29 28 & 0.1328 & 3 & 39812(250)
& 40067(252) & 570(4) & 0.610(\phn61) & 0.91(18) & 11.51 \\
08235+1334 & 08 23 31.9 & +13 34 32 & 0.1368 & 3 & 41011(250)
& 41267(252) & 588(4) & 0.694(\phn69) & 1.05(19) & 11.59 \\
08349+3050 & 08 34 56.5 & +30 50 08 & 0.1085 & 1 & 32523(118)
& 32746(122) & 460(2) & 0.665(\phn66) & 0.65(16) & 11.30 \\
08409+0750 & 08 40 54.2 & +07 50 34 & 0.1029 & 1 & 30862(102)
& 31145(105) & 437(2) & 0.634(\phn57) & $<3.32$ & 11.09--11.57 \\
08433+2702 & 08 43 17.0 & +27 02 34 & 0.1074 & 1 & 32205(116)
& 32447(120) & 456(2) & 0.803(\phn56) & 1.13(17) & 11.42 \\
09049+0137 & 09 04 58.9 & +01 37 12 & 0.1019 & 1 & 30548(224)
& 30860(225) & 433(3) & 0.903(\phn63) & 1.91(19) & 11.50 \\
09116+0334 & 09 11 37.6 & +03 34 27 & 0.1460 & 2 & 43770(\phn70)
& 44085(\phn73) & 631(1) & 1.092(\phn66) & 1.82(20) & 11.86 \\
09302+3241 & 09 30 12.5 & +32 42 00 & 0.1132 & 1 & 33922(\phn68)
& 34177(\phn73) & 482(1) & 0.670(\phn47) & 0.99(13) & 11.40 \\
09425+1751 & 09 42 34.0 & +17 51 49 & 0.1282 & 4 & 38440(\phn41)
& 38750(\phn46) & 550(1) & 0.889(\phn62) & 0.57(13) & 11.54 \\
09517+1458 & 09 51 41.6 & +14 58 47 & 0.1301 & 1 & 39002(127)
& 39324(128) & 559(2) & 0.534(\phn75) & 0.64(13) & 11.40 \\
09525+1602 & 09 52 34.3 & +16 02 09 & 0.1174 & 3 & 35195(250)
& 35515(251) & 502(4) & 0.482(\phn72) & 0.64(13) & 11.27 \\
09540+3521 & 09 54 01.7 & +35 21 17 & 0.1001 & 1 & 30009(300)
& 30266(301) & 424(4) & 0.552(\phn55) & 1.27(13) & 11.28 \\
09576+1858 & 09 57 39.2 & +18 58 53 & 0.1076 & 1 & 32268(106)
& 32583(108) & 458(2) & 0.725(\phn51) & 1.02(14) & 11.38 \\
10034+0726 & 10 03 28.4 & +07 25 47 & 0.1201 & 1 & 36016(133)
& 36358(134) & 514(2) & 0.813(\phn81) & 0.97(15) & 11.51 \\
10040+0932 & 10 04 05.5 & +09 32 00 & 0.1706 & 1 & 51156(104)
& 51495(105) & 746(2) & 0.829(\phn58) & 1.40(15) & 11.89 \\
10086+2621 & 10 08 40.3 & +26 21 37 & 0.1170 & 1 & 35069(108)
& 35366(110) & 499(2) & 0.545(\phn44) & 0.76(14) & 11.33 \\
10113+1736 & 10 11 19.5 & +17 36 52 & 0.1149 & 5 & 34446(\phn\phn\phn)
& 34770(\phn18) & 490(0) & 0.547(\phn49) & 0.92(16) & 11.35 \\
10120+1653 & 10 12 04.6 & +16 53 48 & 0.1247 & 1 & 37386(133)
& 37712(134) & 534(2) & 0.824(\phn58) & 1.66(17) & 11.63 \\
10138+0913 & 10 13 53.9 & +09 13 31 & 0.1023 & 3 & 30669(250)
& 31013(250) & 435(4) & 0.652(\phn52) & 1.13(15) & 11.32 \\
10156+1551 & 10 15 41.0 & +15 51 34 & 0.1110 & 1 & 33290(\phn66)
& 33620(\phn68) & 473(1) & 0.736(\phn59) & 1.52(18) & 11.48 \\
10201+3308 & 10 20 08.4 & +33 08 31 & 0.1256 & 6 & 37654(\phn90)
& 37929(\phn93) & 538(1) & 0.574(\phn57) & 0.69(13) & 11.40 \\
10214+0015 & 10 21 29.1 & +00 15 43 & 0.1252 & 1 & 37527(120)
& 37885(120) & 537(2) & 0.773(\phn62) & 1.00(23) & 11.53 \\
10218+1511 & 10 21 51.9 & +15 11 36 & 0.1105 & 3 & 33127(250)
& 33461(251) & 471(4) & 0.676(\phn54) & 0.92(18) & 11.37 \\
10222+1532 & 10 22 16.9 & +15 32 54 & 0.1096 & 3 & 32857(250)
& 33190(251) & 467(4) & 0.555(\phn50) & 1.05(18) & 11.33 \\
10482+1909 & 10 48 15.4 & +19 09 17 & 0.2187 & 1 & 65576(112)
& 65906(113) & 975(2) & 0.685(\phn75) & 1.44(17) & 12.08 \\
10597+2736 & 10 59 43.0 & +27 36 47 & 0.1276 & 1 & 38255(105)
& 38558(107) & 547(2) & 0.878(\phn62) & 1.59(19) & 11.66 \\
11009+2822 & 11 00 59.7 & +28 22 19 & 0.1309 & 1 & 39243(300)
& 39544(301) & 562(5) & 0.851(\phn60) & $<1.00$ & 11.44--11.60 \\
11119+3257 & 11 11 57.4 & +32 57 49 & 0.1890 & 2 & 56661(\phn70)
& 56944(\phn74) & 831(1) & 1.588(175) & 1.52(17) & 12.19 \\
11175+0917 & 11 17 30.1 & +09 17 59 & 0.1301 & 3 & 39003(250)
& 39357(250) & 559(4) & 0.573(\phn63) & 1.23(22) & 11.53 \\
11188+1138 & 11 18 53.3 & +11 38 51 & 0.1848 & 1 & 55395(107)
& 55744(108) & 812(2) & 0.913(\phn73) & 1.72(22) & 12.03 \\
11233+3451 & 11 23 18.9 & +34 51 33 & 0.1108 & 1 & 33228(119)
& 33502(121) & 472(2) & 0.561(\phn56) & 1.06(14) & 11.34 \\
11243+1655 & 11 24 20.1 & +16 55 51 & 0.1153 & 1 & 34559(197)
& 34895(198) & 492(3) & 0.663(\phn60) & $<1.34$ & 11.22--11.47 \\
11268+1558 & 11 26 47.6 & +15 58 23 & 0.1778 & 1 & 53290(119)
& 53629(120) & 779(2) & 0.720(\phn65) & 1.02(18) & 11.84 \\
11347+2026 & 11 34 42.9 & +20 26 40 & 0.1136 & 3 & 34056(250)
& 34381(251) & 485(4) & 0.756(\phn68) & 1.24(19) & 11.47 \\
11347+2033 & 11 34 45.7 & +20 33 39 & 0.1349 & 1 & 40456(104)
& 40781(105) & 581(2) & 0.790(\phn71) & 1.28(18) & 11.65 \\
11415+0927 & 11 41 35.4 & +09 27 32 & 0.1089 & 1 & 32646(108)
& 32996(109) & 464(2) & 0.847(\phn68) & 1.77(21) & 11.53 \\
11477+2158 & 11 47 46.2 & +21 58 38 & 0.1540 & 1 & 46164(123)
& 46482(124) & 668(2) & 0.604(\phn60) & 0.75(17) & 11.61 \\
11506+1331 & 11 50 39.6 & +13 31 12 & 0.1273 & 7 & 38161(\phn\phn9)
& 38501(\phn16) & 546(0) & 2.583(155) & 3.32(30) & 12.07 \\
11595+1144 & 11 59 32.5 & +11 44 54 & 0.1935 & 1 & 58000(103)
& 58341(104) & 854(2) & 0.945(\phn76) & 1.15(16) & 12.02 \\
12111+2848 & 12 11 08.3 & +28 48 50 & 0.1031 & 8 & 30911(\phn23)
& 31199(\phn31) & 438(0) & 1.194(119) & $<2.84$ & 11.37--11.65 \\
12114+3244 & 12 11 29.3 & +32 44 50 & 0.1055 & 3 & 31636(250)
& 31908(251) & 448(4) & 1.015(122) & 0.90(13) & 11.45 \\
12202+1646 & 12 20 14.7 & +16 46 21 & 0.1810 & 3 & 54262(250)
& 54583(251) & 794(4) & 0.901(\phn81) & 1.24(16) & 11.95 \\
13509+0442 & 13 51 00.2 & +04 42 50 & 0.1360 & 2 & 40772(\phn70)
& 41046(\phn74) & 585(1) & 1.559(\phn94) & 2.53(23) & 11.95 \\
13539+2920 & 13 53 54.4 & +29 20 09 & 0.1085 & 9 & 32513(\phn50)
& 32731(\phn57) & 460(1) & 1.832(128) & 2.73(22) & 11.80 \\
14030+3526 & 14 02 59.0 & +35 26 33 & 0.1079 & 3 & 32349(250)
& 32540(252) & 457(4) & 0.661(\phn66) & 1.09(16) & 11.37 \\
14060+2919 & 14 06 04.3 & +29 19 00 & 0.1168 & 8 & 35009(\phn22)
& 35216(\phn35) & 497(1) & 1.611(161) & 2.42(19) & 11.81 \\
14228+2742 & 14 22 46.2 & +27 42 56 & 0.1380 & 1 & 41372(300)
& 41567(301) & 593(5) & 0.697(\phn98) & 0.99(14) & 11.59 \\
14232+0735 & 14 23 11.8 & +07 35 14 & 0.1547 & 1 & 46377(100)
& 46612(104) & 670(2) & 0.615(\phn74) & 1.18(14) & 11.69 \\
14405+2634 & 14 40 33.0 & +26 34 15 & 0.1074 & 8 & 32212(\phn46)
& 32390(\phn55) & 455(1) & 1.247(\phn62) & 2.19(17) & 11.65 \\
14406+2216 & 14 40 34.6 & +22 16 39 & 0.1108 & 1 & 33216(179)
& 33404(181) & 470(3) & 0.766(\phn77) & 0.84(16) & 11.39 \\
14538+1730 & 14 53 48.9 & +17 30 37 & 0.1035 & 8 & 31041(25)
& 31222(39) & 438(1) & 1.442( 86) & 3.01(30) & 11.71 \\
14575+3256 & 14 57 33.2 & +32 56 51 & 0.1138 & 8 & 34128(\phn22)
& 34271(\phn38) & 483(1) & 1.222(\phn61) & 1.60(14) & 11.64 \\
15001+1433 & 15 00 10.8 & +14 33 15 & 0.1627 & 10 & 48790(\phn70)
& 48968(\phn77) & 706(1) & 1.871(\phn94) & 2.04(20) & 12.13 \\
15005+3555 & 15 00 31.3 & +35 56 06 & 0.1230 & 11 & 36872(\phn24)
& 37004(\phn39) & 524(1) & 0.494(\phn74) & 1.16(12) & 11.42 \\
15059+2835 & 15 05 59.3 & +28 35 37 & 0.1148 & 1 & 34427(116)
& 34570(120) & 488(2) & 0.580(\phn41) & 1.31(13) & 11.42 \\
15158+2747 & 15 15 51.8 & +27 47 01 & 0.1601 & 9 & 48000(100)
& 48133(105) & 693(2) & 0.730(\phn44) & 1.67(17) & 11.83 \\
15168+0045 & 15 16 50.6 & +00 45 51 & 0.1539 & 12 & 46138(300)
& 46310(302) & 665(5) & 0.610(\phn73) & 0.91(16) & 11.64 \\
15206+3631 & 15 20 38.4 & +36 31 37 & 0.1524 & 11 & 45680(\phn47)
& 45788(\phn57) & 657(1) & 0.771(\phn46) & 1.06(12) & 11.72 \\
15206+3342 & 15 20 38.6 & +33 42 12 & 0.1244 & 8 & 37297(\phn63)
& 37411(\phn71) & 530(1) & 1.743(139) & 1.95(18) & 11.86 \\
15225+2350 & 15 22 32.9 & +23 50 36 & 0.1390 & 2 & 41671(\phn70)
& 41803(\phn77) & 596(1) & 1.300(\phn91) & 1.48(16) & 11.83 \\
16075+2838 & 16 07 31.0 & +28 38 45 & 0.1697 & 9 & 50861(\phn50)
& 50925(\phn60) & 737(1) & 0.841(\phn67) & 1.01(17) & 11.83 \\
16122+3528 & 16 12 16.5 & +35 28 27 & 0.1245 & 1 & 37314(145)
& 37362(149) & 529(2) & 0.778(\phn62) & 1.02(13) & 11.53 \\
16283+0442 & 16 28 22.8 & +04 42 17 & 0.1235 & 1 & 37029(104)
& 37090(110) & 525(2) & 0.597(\phn48) & 0.95(24) & 11.43 \\
16336+1019 & 16 33 38.0 & +10 19 50 & 0.1455 & 1 & 43616(106)
& 43664(112) & 625(2) & 0.575(\phn63) & $<1.85$ & 11.36--11.71 \\
16380+1508 & 16 38 03.1 & +15 08 00 & 0.1623 & 1 & 48644(171)
& 48681(175) & 702(3) & 0.541(\phn76) & $<1.23$ & 11.43--11.71 \\
16544+3212 & 16 54 31.0 & +32 12 36 & 0.1363 & 1 & 40876(119)
& 40872(124) & 582(2) & 0.619(\phn62) & 1.04(18) & 11.55 \\
17023+0232 & 17 02 21.0 & +02 32 41 & 0.1385 & 1 & 41511(117)
& 41520(123) & 592(2) & 0.587(\phn47) & $<1.42$ & 11.32--11.61 \\
17114+2059 & 17 11 27.0 & +20 59 29 & 0.1210 & 1 & 36268(115)
& 36249(120) & 513(2) & 1.013(\phn51) & 1.17(25) & 11.59 \\
17129+1004 & 17 12 58.9 & +10 04 13 & 0.1130 & 1 & 33877(\phn\phn\phn)
& 33864(\phn37) & 477(1) & 0.580(\phn52) & 1.39(40) & 11.42 \\
20090+0129 & 20 09 03.8 & +01 29 05 & 0.1015 & 1 & 30437(106)
& 30178(109) & 423(2) & 0.665(\phn53) & $<3.87$ & 11.08--11.60 \\
20246+0106 & 20 24 40.4 & +01 06 29 & 0.1149 & 8 & 34440(\phn37)
& 34164(\phn45) & 481(1) & 1.180(\phn83) & $<2.64$ & 11.45--11.72 \\
21477+0502 & 21 47 45.9 & +05 02 03 & 0.1710 & 2 & 51265(\phn70)
& 50919(\phn71) & 737(1) & 1.139(148) & 1.46(22) & 11.98 \\
21534+3504 & 21 53 26.4 & +35 04 40 & 0.1038 & 8 & 31111(293)
& 30802(294) & 432(4) & 0.972(156) & 4.47(40) & 11.71 \\
22139+2448 & 22 13 54.0 & +24 47 51 & 0.1534 & 1 & 45981(123)
& 45640(124) & 655(2) & 0.670(\phn60) & 1.60(29) & 11.75 \\
22285+3555 & 22 28 32.8 & +35 55 24 & 0.1175 & 8 & 35229(\phn23)
& 34911(\phn29) & 493(0) & 1.218(134) & 1.65(20) & 11.66 \\
22368+0904 & 22 36 52.8 & +09 04 55 & 0.1080 & 1 & 32376(104)
& 32011(104) & 450(2) & 0.875(105) & $<2.58$ & 11.26--11.59 \\
22583+1703 & 22 58 21.4 & +17 03 25 & 0.1191 & 1 & 35713(108)
& 35350(108) & 499(2) & 0.606(\phn61) & $<2.30$ & 11.19--11.58 \\
22584+2348 & 22 58 21.5 & +23 48 16 & 0.1024 & 1 & 30685(121)
& 30333(121) & 425(2) & 0.587(\phn70) & $<2.67$ & 11.03--11.48 \\
23018+0333 & 23 01 50.2 & +03 33 45 & 0.1185 & 1 & 35527(181)
& 35159(181) & 496(3) & 0.620(\phn87) & 1.10(18) & 11.42 \\
23055+2127 & 23 05 35.4 & +21 27 14 & 0.1021 & 1 & 30596(102)
& 30239(102) & 424(2) & 0.689(\phn69) & $<1.75$ & 11.10--11.40 \\
23068+3014 & 23 06 55.3 & +30 14 14 & 0.1313 & 3 & 39362(250)
& 39023(250) & 554(4) & 0.704(\phn63) & $<1.84$ & 11.34--11.65 \\
23073+0005 & 23 07 21.6 & +00 05 39 & 0.1037 & 3 & 31088(250)
& 30722(250) & 431(4) & 0.789(\phn63) & 1.64(18) & 11.43 \\
23233+2817 & 23 23 20.7 & +28 17 47 & 0.1140 & 8 & 34179(\phn37)
& 33836(\phn39) & 477(1) & 1.262(126) & 2.11(34) & 11.68 \\
23327+2913 & 23 32 42.7 & +29 13 25 & 0.1067 & 10 & 31981(\phn22)
& 31641(\phn26) & 444(0) & 2.099(126) & 2.81(45) & 11.81 \\
23498+2423 & 23 49 52.4 & +24 23 28 & 0.2120 & 2 & 63556(\phn70)
& 63209(\phn71) & 932(1) & 1.025(\phn82) & 1.45(30) & 12.15 \\
23580+2636 & 23 58 05.3 & +26 36 11 & 0.1439 & 1 & 43145(423)
& 42805(423) & 611(6) & 0.634(\phn57) & 1.79(21) & 11.70 \\
\enddata
\tablerefs{Redshifts were obtained from:  (1) \citet{sau00}; (2) \citet{kim98}; 
(3) \citet{law99}; (4) \citet{shu98}; (5) \citet{mor96b};  (6) \citet{deg92};
(7) \citet{dow93}; (8) \citet{fis95};  (9) \citet{str88};  (10) \citet{str92}; 
(11) \citet{dey90}; (12) \citet{lee94}; (13) \citet{all85}. }
\end{deluxetable}

\begin{deluxetable}{cccccrrrrc} 
\tabletypesize{\scriptsize}
\tablecaption{OH Non-Detections:  OH Limits and 1.4 GHz Properties\label{nondetectOH}}
\tablewidth{0pt}
\tablehead{
\colhead{{\it IRAS} Name} &  \colhead{$z_\odot$}  
& \colhead{$\log L_{FIR}$} & \colhead{$\log L^{pred}_{OH}$} 
& \colhead{$\log L^{max}_{OH}$} & \colhead{$t_{on}$} & \colhead{RMS} 
& \colhead{$f_{1.4GHz}$\tablenotemark{a}} & \colhead{Class\tablenotemark{b}} 
& \colhead{Note} \\
\colhead{FSC} &  & \colhead{$h^{-2}_{75} L_\odot$} &
\colhead{$h^{-2}_{75} L_\odot$} & \colhead{$h^{-2}_{75} L_\odot$} & \colhead{min} 
& \colhead{mJy} & \colhead{mJy} & \\
\colhead{(1)}& \colhead{(2)}& \colhead{(3)}& \colhead{(4)}& \colhead{(5)}& 
\colhead{(6)}& \colhead{(7)}& \colhead{(8)}& \colhead{(9)}& \colhead{(10)}
 }
\startdata
00051+2657 & 0.1254 & 11.58 & 1.96 & 1.76 & 12 & 0.59 & 5.4(0.5) & \\
03248+1756 & 0.1257 & 11.52 & 1.88 & 1.81 & 12 & 0.65 & 4.2(0.5) & & 2\\
03250+1606 & 0.1290 & 11.80 & 2.26 & 1.97 &\phn8& 0.90 & 9.8(0.6) & L(1) & 2\\
03477+2611 & 0.1494 & 11.71 & 2.15 & 1.92 & 12 & 0.59 & 4.1(0.5) & & 6\\
04229+0056 & 0.1530 & 11.34--11.71 & 1.63--2.14 & 1.92 & 16 & 0.57 & 6.5(0.5) & & 1\\
04479+0616 & 0.1158 & 11.65 & 2.06 & 1.46 & 32 & 0.34 & 6.6(0.5) & \\
06268+3509 & 0.1698 & 11.88 & 2.37 & 2.07 & 12 & 0.64 & 5.3(0.5) & \\
06368+2812 & 0.1249 & 11.76 & 2.20 & 1.69 & 16 & 0.49 & 12.1(0.6)& \\
06561+1902 & 0.1882 & 12.02 & 2.57 & 2.12 & 12 & 0.59 & 4.9(0.5) & \\
07178+1952 & 0.1148 & 11.48 & 1.82 & 1.67 & 12 & 0.56 & 3.7(0.6) & \\
07188+0407 & 0.1035 & 11.09--11.37 & 1.28--1.68 & 1.57 & 12 & 0.55 & 18.1(1.0) & & 1\\
07241+3052 & 0.1112 & 11.37 & 1.67 & 1.62 & 12 & 0.53 & 4.0(0.5) & \\
07328+0457 & 0.1300 & 11.67 & 2.09 & 1.82 & 12 & 0.61 & 6.4(0.5) & & 2 \\
07381+3215 & 0.1703 & 11.75 & 2.19 & 1.94 & 24 & 0.47 & 4.3(0.5) & & 2,4\\
08003+0734 & 0.1179 & 11.52 & 1.88 & 1.71 & 12 & 0.57 & 6.0(0.5) & \\
08007+0711 & 0.1405 & 11.69 & 2.12 & 1.79 & 12 & 0.49 & 11.7(0.6)& \\
08012+0125 & 0.2203 & 11.99 & 2.53 & 2.24 & 12 & 0.56 & $<5.0$\phn & & 5\\
08122+0505 & 0.1030 & 11.55 & 1.92 & 1.61 & 12 & 0.60 & 8.0(0.5) & \\
08132+1628 & 0.1004 & 11.43 & 1.75 & 1.59 & 12 & 0.61 & 9.1(0.6) & \\
08147+3137 & 0.1239 & 11.27--11.39 & 1.53--1.70 & 1.80 & 12 & 0.65 & 73.2(2.6)& A(6)& 1 \\
08200+1931 & 0.1694 & 11.84 & 2.32 & 2.05 & 12 & 0.62 & $<5.0$\phn & & 4 \\
08224+1329 & 0.1328 & 11.51 & 1.86 & 1.79 & 12 & 0.55 & 3.9(0.5) & \\
08235+1334 & 0.1368 & 11.59 & 1.97 & 1.79 & 12 & 0.51 & $<5.0$\phn & \\
08349+3050 & 0.1085 & 11.30 & 1.57 & 1.66 & 12 & 0.61 & 2.7(0.5) & & 1\\
08409+0750 & 0.1029 & 11.09--11.57 & 1.29--1.95 & 1.60 & 12 & 0.59 & $<5.0$\phn & & 1 \\
08433+2702 & 0.1074 & 11.42 & 1.74 & 1.65 & 12 & 0.61 & 3.8(0.6) & \\
09049+0137 & 0.1019 & 11.50 & 1.85 & 1.71 & 12 & 0.77 & 13.2(0.6) & \\
09116+0334 & 0.1460 & 11.86 & 2.35 & 1.97 & 12 & 0.69 & 11.0(0.6) & L(1) \\
09302+3241 & 0.1132 & 11.40 & 1.71 & 1.83 & 12 & 0.83 & 5.1(0.5) & & 1 \\
09425+1751 & 0.1282 & 11.54 & 1.90 & 1.82 & 12 & 0.63 &46.1(1.5) & S2(2)\\
09517+1458 & 0.1301 & 11.40 & 1.71 & 1.85 & 12 & 0.65 &39.0(1.3) & & 1 \\
09525+1602 & 0.1174 & 11.27 & 1.54 & 1.68 & 12 & 0.55 & $<5.0$\phn & & 1 \\
09540+3521 & 0.1001 & 11.28 & 1.55 & 1.59 & 12 & 0.60 & 8.0(0.5) & & 1 \\
09576+1858 & 0.1076 & 11.38 & 1.69 & 1.60 & 16 & 0.54 & 8.4(1.0) & \\
10034+0726 & 0.1201 & 11.51 & 1.86 & 1.83 &\phn8& 0.73 & $<5.0$\phn & &  \\
10040+0932 & 0.1706 & 11.89 & 2.39 & 2.09 & 12 & 0.66 & 3.5(0.6) & & 4 \\
10086+2621 & 0.1170 & 11.33 & 1.62 & 1.72 & 12 & 0.60 & 5.6(0.6) & & 1 \\
10113+1736 & 0.1149 & 11.35 & 1.64 & 1.73 & 12 & 0.63 & 4.5(0.6) & C(5) & 1 \\
10120+1653 & 0.1247 & 11.63 & 2.03 & 1.54 & 44 & 0.35 & 10.0(0.5)& &   \\
10138+0913 & 0.1023 & 11.32 & 1.61 & 1.63 & 12 & 0.63 & 2.9(0.5) & & 1 \\
10156+1551 & 0.1110 & 11.48 & 1.82 & 1.68 & 12 & 0.61 & 7.3(1.6) & &   \\
10201+3308 & 0.1256 & 11.40 & 1.71 & 1.79 & 12 & 0.60 & 3.8(0.5) & H(4) & 1 \\
10214+0015 & 0.1252 & 11.53 & 1.90 & 1.84 & 12 & 0.69 & 7.1(0.6) & \\
10218+1511 & 0.1105 & 11.37 & 1.67 & 1.69 & 12 & 0.63 & 4.2(0.5) & & 1 \\
10222+1532 & 0.1096 & 11.33 & 1.62 & 1.67 & 12 & 0.60 & 4.9(0.5) & & 1 \\
10482+1909 & 0.2187 & 12.08 & 2.65 & 2.23 & 12 & 0.56 & 9.0(1.4) & & 2 \\
10597+2736 & 0.1276 & 11.66 & 2.07 & 1.73 & 12 & 0.51 & 5.8(0.5) & \\
11009+2822 & 0.1309 & 11.44--11.60 & 1.77--1.99 & 1.77 & 12 & 0.54 & $<5.0$\phn & &  \\
11119+3257 & 0.1890 & 12.19 & 2.80 & 2.05 & 24 & 0.49 & 110.4(3.3) & S1(1) &  \\
11175+0917 & 0.1301 & 11.53 & 1.88 & 1.79 & 12 & 0.57 & 5.5(0.5) & \\
11188+1138 & 0.1848 & 12.03 & 2.58 & 2.13 & 12 & 0.61 & 7.5(0.5) & & 3 \\
11233+3451 & 0.1108 & 11.34 & 1.64 & 1.68 & 12 & 0.61 & $<5.0$\phn & & 1 \\
11243+1655 & 0.1153 & 11.22--11.47 & 1.46--1.80 & 1.68 & 12 & 0.56 & 3.7(0.5) & & 1 \\
11268+1558 & 0.1778 & 11.84 & 2.32 & 2.00 & 12 & 0.49 & 3.3(0.5) & & 4 \\
11347+2026 & 0.1136 & 11.47 & 1.81 & 1.68 & 12 & 0.57 & 4.1(0.6) & & \\
11347+2033 & 0.1349 & 11.65 & 2.05 & 1.77 & 12 & 0.50 & 8.3(0.6) & & \\
11415+0927 & 0.1089 & 11.53 & 1.89 & 1.66 & 12 & 0.60 & 8.4(0.5) & \\
11477+2158 & 0.1540 & 11.61 & 2.00 & 1.93 & 12 & 0.56 & $<5.0$\phn & \\
11506+1331 & 0.1273 & 12.07 & 2.64 & 1.91 & 12 & 0.79 &13.9(0.6) & H(1) & 3 \\
11595+1144 & 0.1935 & 12.02 & 2.56 & 2.23 & 12 & 0.71 & 9.4(1.0) & & 3 \\
12111+2848 & 0.1031 & 11.37--11.65 & 1.67--2.06 & 1.76 &\phn8& 0.85 & 7.3(0.5) &  & 1,3 \\
12114+3244 & 0.1055 & 11.45 & 1.78 & 1.67 & 12 & 0.65 & 4.6(0.5) & & 3 \\
12202+1646 & 0.1810 & 11.95 & 2.47 & 2.15 &  8 & 0.68 & 9.4(0.6) & S2(7) & 4 \\
13509+0442 & 0.1360 & 11.95 & 2.47 & 2.07 &  4 & 0.99 & 10.3(0.6)& H(1) &  \\
13539+2920 & 0.1085 & 11.80 & 2.26 & 1.67 & 12 & 0.62 & 12.1(0.6)& H(1) &  \\
14030+3526 & 0.1079 & 11.37 & 1.66 & 1.73 & 12 & 0.74 & 3.6(0.5) & & 1  \\
14060+2919 & 0.1168 & 11.81 & 2.28 & 1.65 & 16 & 0.51 & 9.7(0.5) & H(1) & \\
14228+2742 & 0.1380 & 11.59 & 1.97 & 1.78 & 12 & 0.50 & $<5.0$\phn & \\
14232+0735 & 0.1547 & 11.69 & 2.11 & 1.98 & 12 & 0.62 & 2.7(0.6) & \\
14405+2634 & 0.1074 & 11.65 & 2.05 & 1.65 & 12 & 0.61 & 6.5(0.5) & \\
14406+2216 & 0.1108 & 11.39 & 1.70 & 1.72 & 12 & 0.67 & $<5.0$\phn & & 1 \\
14538+1730 & 0.1035 & 11.71 & 2.14 & 1.67 & 12 & 0.68 & 13.3(0.9) & \\
14575+3256 & 0.1138 & 11.64 & 2.05 & 1.80 &  8 & 0.76 & 6.8(0.5) & \\
15001+1433 & 0.1627 & 12.13 & 2.72 & 2.01 & 12 & 0.61 & 16.9(1.0)& S2(1) &\\
15005+3555 & 0.1230 & 11.42 & 1.74 & 1.78 & 12 & 0.63 & $<5.0$\phn & & 1 \\
15059+2835 & 0.1148 & 11.42 & 1.74 & 1.67 & 12 & 0.56 & 3.4(0.5) & \\
15158+2747 & 0.1601 & 11.83 & 2.31 & 1.94 & 12 & 0.53 & 7.4(0.5) & & 4 \\
15168+0045 & 0.1539 & 11.64 & 2.04 & 2.09 &  8 & 0.82 & 14.6(0.6)& & 1 \\
15206+3631 & 0.1524 & 11.72 & 2.15 & 1.95 & 12 & 0.61 & 5.0(0.5) & \\
15206+3342 & 0.1244 & 11.86 & 2.34 & 1.72 & 12 & 0.54 & 11.2(0.6)& H(1) & \\
15225+2350 & 0.1390 & 11.83 & 2.31 & 1.86 & 12 & 0.60 & 6.9(0.5) & H(1) & \\
16075+2838 & 0.1697 & 11.83 & 2.31 & 2.02 & 16 & 0.58 & 4.6(0.5) & A(3) & 4 \\
16122+3528 & 0.1245 & 11.53 & 1.89 & 1.89 & 12 & 0.79 & 4.2(0.6) &  &  \\
16283+0442 & 0.1235 & 11.43 & 1.76 & 1.72 & 12 & 0.55 & 6.1(0.5) & \\
16336+1019 & 0.1455 & 11.36--11.71 & 1.66--2.14 & 1.98 & 12 & 0.72 & 3.4(0.6) & & 1,3 \\
16380+1508 & 0.1623 & 11.43--11.71 & 1.76--2.14 & 1.67 & 12 & 0.58 & $<5.0$\phn & \\
16544+3212 & 0.1363 & 11.55 & 1.92 & 2.03 &  8 & 0.92 & 4.5(0.5) & & 1 \\
17023+0232 & 0.1385 & 11.32--11.61 & 1.61--2.00 & 1.89 & 12 & 0.64 & 5.5(0.5) && 1 \\
17114+2059 & 0.1210 & 11.59 & 1.98 & 1.87 & 12 & 0.80 & 8.8(0.5) & & 3 \\
17129+1004 & 0.1130 & 11.42 & 1.73 & 1.74 & 12 & 0.68 & 3.8(0.7) & \\
20090+0129 & 0.1015 & 11.08--11.60 & 1.28--1.98 & 1.55 & 16 & 0.55 & 12.4(1.0)&& 1 \\
20246+0106 & 0.1149 & 11.45--11.72 & 1.78--2.15 & 1.76 & 12 & 0.71 & 6.4(0.5) & \\
21477+0502 & 0.1710 & 11.98 & 2.51 & 2.04 & 12 & 0.61 & 6.9(0.5) & L(1) & 4 \\
21534+3504 & 0.1038 & 11.71 & 2.14 & 1.66 & 12 & 0.69 & $<5.0$\phn & \\
22139+2448 & 0.1534 & 11.75 & 2.20 & 1.88 & 12 & 0.52 & 3.9(0.5) & \\
22285+3555 & 0.1175 & 11.66 & 2.07 & 1.72 & 16 & 0.62 & 6.7(0.5) & \\
22368+0904 & 0.1080 & 11.26--11.59 & 1.51--1.97 & 1.67 & 12 & 0.65 & 4.2(0.5) && 1,2 \\
22583+1703 & 0.1191 & 11.19--11.58 & 1.42--1.96 & 1.69 & 12 & 0.56 & 12.6(1.5)&& 1 \\
22584+2348 & 0.1024 & 11.03--11.48 & 1.21--1.82 & 1.60 & 12 & 0.61 & $<5.0$\phn & & 1\\
23018+0333 & 0.1185 & 11.42 & 1.74 & 1.68 & 12 & 0.55 & $<5.0$\phn & \\
23055+2127 & 0.1021 & 11.10--11.40 & 1.30--1.71 & 1.61 & 12 & 0.64 & 4.0(0.5) & & 1 \\
23068+3014 & 0.1313 & 11.34--11.65 & 1.63--2.05 & 1.72 & 16 & 0.50 & $<5.0$\phn & & 1 \\
23073+0005 & 0.1037 & 11.43 & 1.75 & 1.60 & 12 & 0.61 & 5.5(0.6) & \\
23233+2817 & 0.1140 & 11.68 & 2.10 & 1.65 & 12 & 0.55 & 35.5(1.2)& S2(1) & \\
23327+2913 & 0.1067 & 11.81 & 2.27 & 1.68 & 16 & 0.68 & 8.4(0.6) & L(1) & 2 \\
23498+2423 & 0.2120 & 12.15 & 2.74 & 2.22 & 12 & 0.59 & 6.8(0.5) & S2(1) & \\
23580+2636 & 0.1439 & 11.70 & 2.13 & 1.85 & 12 & 0.55 & $<5.0$\phn & &\\
\enddata
\tablenotetext{a}{1.4 GHz continuum fluxes are courtesy of the NRAO VLA Sky Survey 
\citep{con98}.}
\tablenotetext{b}{Spectral classifications use the codes:  
``S2'' = Seyfert type 2;  ``S1'' = Seyfert type 1;  ``A'' = Active;  
``C'' = Composite AGN and starburst;  ``H'' = HII region (starburst);  
and ``L'' = low-ionization emission region (LINER).  }
\tablerefs{Spectral classifications were obtained from: (1) Veilleux, 
Kim, \& Sanders (1999); (2) \citet{vei95};  
(3) \citet{str88}; (4) \citet{deg92}; (5) \citet{mor96b}; (6) \citet{con95}; (7) \citet{law99}.}
\tablecomments{(1) Source needs more integration time, based on
$L^{pred}_{OH} < L^{max}_{OH}$; (2) Source needs more integration time, due
to a suggestive feature in the bandpass; 
(3) Observations were performed during daylight, which increases the 
RMS noise significantly; (4) Galactic HI in bandpass; (5) RFI in bandpass --- there is a 
small chance that the OH line falls on one of the two narrow RFI regions and is itself narrow;
(6) Re-observation of a nondetection listed in Paper I.}
\end{deluxetable}

\begin{deluxetable}{rccccrrrrcc}
\tabletypesize{\scriptsize}
\tablecaption{OH Detections:  Optical Redshifts and FIR Properties 
		\label{detectFIR}}
\tablewidth{0pt}
\tablehead{
\colhead{{\it IRAS} Name} &  \colhead{$\alpha$} & \colhead{$\delta$} & 
\colhead{$z_\odot$} & 
\colhead{Ref} &
\colhead{$v_\odot$} & 
\colhead{$v_{CMB}$} & 
\colhead{$D_L$} &
\colhead{$f_{60\mu m}$} &
\colhead{$f_{100\mu m}$} &
\colhead{$\log L_{FIR}$} 
 \\
\colhead{FSC} & \colhead{B1950} & \colhead{B1950} &\colhead{} &
\colhead{z} & \colhead{km/s} & \colhead{km/s} & \colhead{$h^{-1}_{75}$Mpc} & 
\colhead{Jy} & \colhead{Jy} & \colhead{$h^{-2}_{75} L_\odot$} \\
\colhead{(1)}&\colhead{(2)}&\colhead{(3)}&\colhead{(4)}&\colhead{(5)}&
\colhead{(6)}&\colhead{(7)}&\colhead{(8)}&\colhead{(9)}&\colhead{(10)}&
\colhead{(11)}
}
\startdata
04121+0223 & 04 12 10.5 & +02 23 12 & 0.1216 & 3 & 36454(250)
& 36362(253) & 514(4) & 0.889(\phn62) & $<2.15$ & 11.38--11.67 \\
07163+0817 & 07 16 23.7 & +08 17 34 & 0.1107 & 1 & 33183(110)
& 33367(115) & 470(2) & 0.891(\phn89) & 1.37(11) & 11.51 \\
07572+0533 & 07 57 17.9 & +05 33 16 & 0.1894 & 1 & 56783(122)
& 57022(126) & 833(2) & 0.955(\phn76) & 1.30(20) & 12.01 \\
08201+2801 & 08 20 10.1 & +28 01 19 & 0.1680 & 2 & 50365(\phn70)
& 50583(\phn77) & 731(1) & 1.171(\phn70) & 1.43(16) & 11.97 \\
08279+0956 & 08 27 56.1 & +09 56 41 & 0.2085 & 1 & 62521(107)
& 62788(110) & 925(2) & 0.586(\phn64) & $<1.26$ & 11.71--11.97 \\
08449+2332 & 08 44 55.6 & +23 32 12 & 0.1510 & 1 & 45277(102)
& 45530(106) & 653(2) & 0.867(\phn69) & 1.20(17) & 11.76 \\
08474+1813 & 08 47 28.3 & +18 13 14 & 0.1450 & 2 & 43470(\phn70)
& 43739(\phn75) & 626(1) & 1.279(115) & 1.54(18) & 11.88 \\
09039+0503 & 09 03 56.4 & +05 03 28 & 0.1250 & 2 & 37474(\phn70)
& 37781(\phn73) & 535(1) & 1.484(\phn89) & 2.06(21) & 11.83 \\
09531+1430 & 09 53 08.3 & +14 30 22 & 0.2151 & 1 & 64494(148)
& 64818(149) & 958(2) & 0.777(\phn62) & 1.04(14) & 12.04 \\
09539+0857 & 09 53 54.9 & +08 57 23 & 0.1290 & 2 & 38673(\phn70)
& 39008(\phn72) & 554(1) & 1.438(101) & 1.04(18) & 11.76 \\
10339+1548 & 10 33 58.1 & +15 48 11 & 0.1965 & 1 & 58906(122)
& 59242(123) & 868(2) & 0.977(\phn59) & 1.35(16) & 12.06 \\
11028+3130 & 11 02 54.0 & +31 30 40 & 0.1990 & 2 & 59659(\phn70)
& 59948(\phn73) & 879(1) & 1.021(\phn72) & 1.44(16) & 12.10 \\
11524+1058 & 11 52 29.6 & +10 58 22 & 0.1784 & 1 & 53479(134)
& 53823(135) & 782(2) & 0.821(\phn66) & 1.17(15) & 11.90 \\
12032+1707 & 12 03 14.9 & +17 07 48 & 0.2170 & 2 & 65055(\phn70)
& 65382(\phn72) & 967(1) & 1.358(\phn95) & 1.54(19) & 12.27 \\
15224+1033 & 15 22 27.4 & +10 33 17 & 0.1348 & 1 & 40405(155)
& 40559(158) & 577(2) & 0.737(\phn74) & 0.72(15) & 11.54 \\
15587+1609 & 15 58 45.5 & +16 09 23 & 0.1375 & 1 & 41235(195)
& 41329(198) & 589(3) & 0.740(\phn52) & 0.82(21) & 11.57 \\
16100+2527 & 16 10 00.4 & +25 28 02 & 0.1310 & 3 & 39272(250)
& 39338(252) & 559(4) & 0.715(\phn50) & $<1.38$ & 11.36--11.60 \\
16255+2801 & 16 25 34.0 & +28 01 32 & 0.1340 & 1 & 40186(122)
& 40226(127) & 572(2) & 0.885(\phn88) & 1.26(26) & 11.66 \\
22055+3024 & 22 05 33.6 & +30 24 52 & 0.1269 & 1 & 38041(\phn24)
& 37715(\phn29) & 534(0) & 1.874(356) & 2.32(23) & 11.91 \\
23019+3405 & 23 01 57.3 & +34 05 27 & 0.1080 & 4 & 32389(\phn28)
& 32061(\phn32) & 450(0) & 1.417(\phn99) & 2.11(38) & 11.67 \\
23028+0725 & 23 02 49.2 & +07 25 35 & 0.1496 & 1 & 44845(198)
& 44476(198) & 637(3) & 0.914(100) & $<1.37$ & 11.58--11.78 \\
23129+2548 & 23 12 54.4 & +25 48 13 & 0.1790 & 2 & 53663(\phn70)
& 53314(\phn71) & 774(1) & 1.811(145) & 1.64(44) & 12.18 \\
23199+0123 & 23 19 57.7 & +01 22 57 & 0.1367 & 3 & 40981(250)
& 40614(250) & 578(4) & 0.627(\phn63) & 1.03(16) & 11.55 \\
23234+0946 & 23 23 23.6 & +09 46 15 & 0.1279 & 4 & 38356(\phn24)
& 37988(\phn24) & 539(0) & 1.561(\phn94) & 2.11(30) & 11.85 \\
\enddata
\tablerefs{Redshifts were obtained from:  (1) \citet{sau00}; (2) \citet{kim98}; 
(3) \citet{law99}; (4) \citet{fis95}.}
\end{deluxetable}

\begin{deluxetable}{cccrcccrcccr}
\tabletypesize{\scriptsize}
\tablecaption{OH Detections:  OH Line and 1.4 GHz Continuum Properties \label{detectOH}}
\tablewidth{0pt}
\tablehead{
\colhead{{\it IRAS} Name} &  
\colhead{$v_{1667,\odot}$} & 
\colhead{$t_{on}$} & 
\colhead{$f_{1667}$} & 
\colhead{$W_{1667}$} &
\colhead{$\Delta \nu_{1667}$\tablenotemark{a}} & 
\colhead{$\Delta v_{1667}$\tablenotemark{b}} & 
\colhead{$R_H$} & 
\colhead{$\log L_{FIR}$} & 
\colhead{$\log L^{pred}_{OH}$} & 
\colhead{$\log L_{OH}$} &
\colhead{$f_{1.4GHz}$\tablenotemark{c}}  \\
\colhead{FSC} &\colhead{km/s} & \colhead{min} & \colhead{mJy} & \colhead{MHz}
& \colhead{MHz} & \colhead{km/s} &\colhead{} & \colhead{$h^{-2}_{75} L_\odot$} 
& \colhead{$h^{-2}_{75} L_\odot$} & \colhead{$h^{-2}_{75} L_\odot$} & \colhead{mJy} \\
\colhead{(1)}&\colhead{(2)}&\colhead{(3)}&\colhead{(4)}&\colhead{(5)}&
\colhead{(6)}&\colhead{(7)}&\colhead{(8)}&\colhead{(9)}&\colhead{(10)}&
\colhead{(11)}&\colhead{(12)}
}
\startdata
04121+0223 & 36590(14) & 56 &  2.52 & 0.76 & 1.04 & 209 & 2.9 & 11.38--11.67 & 1.69--2.08 & 2.30 & 3.1(0.5)\\
07163+0817 & 33150(14) & 80 &  4.00 & 0.69 & 0.12 & \phn24 &$\sim\phn5.5$& 11.51 & 1.86 & 2.35 & 3.5(0.5)\\
07572+0533 & 56845(15) & 72 &  2.26 & 1.03 & 0.73 & 156 &$\geq10.4$ & 12.01 & 2.56 & 2.71 & $< 5.0$\phn\\
08201+2801 & 50325(15) & 20 & 14.67 & 0.97--1.19 & 0.98 & 205 &$\geq\phn8.2$& 11.97 & 2.50 & 3.42 & 16.7(0.7)\\
08279+0956 & 62422(15) & 20 &  4.79 & 1.02 & 0.95 & 207 & 5.9 & 11.71--11.97 & 2.14--2.50 & 3.19 & 4.4(0.8)\\
08449+2332 & 45424(14) & 40 &  2.49 & 1.09 & 0.47 & \phn97 &$\geq11.0$& 11.76 & 2.21 & 2.56 & 6.1(0.5)\\
08474+1813 & 43750(14) & 36 &  2.20 & 1.29--1.70 & 1.98 & 409 &$\geq\phn3.0$& 11.88 & 2.37 & 2.67 & 4.2(0.5)\\
09039+0503 & 37720(14) & 48 &  5.17 & 1.23 & 1.05 & 212 & 8.5 & 11.83 & 2.30 & 2.80 & 6.6(0.5)\\
09531+1430 & 64434(15) & 40 &  3.98 & 1.03 & 1.17 & 256 &$\sim\phn3.4$& 12.04 & 2.60 & 3.38 & 3.0(0.5)\\
09539+0857 & 38455(14) & 36 & 14.32 & 1.47 & 1.56 & 317 & 2.5 & 11.76 & 2.21 & 3.45 & 9.5(1.2)\\
10339+1548 & 58983(15) & 28 &  6.26 & 0.28 & 0.19 &\phn40&$\gtrsim14.5$& 12.06 & 2.63 & 2.62 & 5.1(0.5)\\
11028+3130 & 59619(15) & 28 &  4.27 & 0.72 & 0.41 &\phn89& 5.5 & 12.10 & 2.67 & 2.94 & $< 5.0$\phn\\
11524+1058 & 53404(15) & 40 &  3.17 & 1.21 & 1.32 & 279 &$\sim\phn4.9$& 11.90 & 2.40 & 2.95 & $< 5.0$\phn\\
12032+1707 & 64920(15) & 32 & 16.27 & 2.69 & 3.90 & 853 &\nodata& 12.27 & 2.91 & 4.11 & 28.7(1.0)\\
15224+1033 & 40290(14) & 32 & 12.27 & 0.73--0.80 & 0.15 &\phn31&$\geq\phn9.5$& 11.54 & 1.90 & 3.01 & 3.6(0.5)\\
15587+1609 & 40938(14) & 24 & 13.91 & 0.99 & 0.86 & 176 & 6.9 & 11.57 & 1.95 & 3.23 & $< 5.0$\phn\\
16100+2527 & 40040(14) & 72 &  2.37 & 0.60 & 0.23 &\phn46& 3.2 & 11.36--11.60 & 1.65--1.99 & 2.26 & $< 5.0$\phn\\
16255+2801 & 40076(14) & 40 &  7.02 & 0.45 & 0.39 &\phn79&$\geq13.7$& 11.66 & 2.07 & 2.54 & $< 5.0$\phn\\
22055+3024 & 37965(14) & 48 &  6.35 & 0.77 & 0.46 &\phn92& 6.2 & 11.91 & 2.41 & 2.71 & 6.4(0.5)\\
23019+3405 & 32294(14) & 32 &  3.58 & 0.52 & 0.28 &\phn57&$\geq15.6$& 11.67 & 2.08 & 2.10 & 7.7(0.5)\\
23028+0725 & 44529(14) & 28 &  8.69 & 1.09 & 1.06 & 219 & 1.9 & 11.58--11.78 & 1.96--2.23 & 3.26 & 19.5(1.1)\\
23129+2548 & 53394(15) & 32 &  4.59 & 2.0 & 1.78 & 376 &\nodata& 12.18 & 2.78 & 3.24 & 4.7(0.5)\\
23199+0123 & 40680(14) & 52 &  1.80 & 0.82 & 0.68 & 139 &$\sim\phn2.3$& 11.55 & 1.91 & 2.35 & 3.0(0.5)\\
23234+0946 & 38240(14) & 24 &  3.32 & 1.23 & 1.32 & 266 & 2.4 & 11.85 & 2.33 & 2.72 & 11.6(1.0)\\
\enddata
\tablenotetext{a}{$\Delta \nu_{1667}$ is the {\it observed} FWHM.}
\tablenotetext{b}{$\Delta v_{1667}$ is the {\it rest frame} FWHM.  The rest
frame and observed widths are related by 
$\Delta v_{rest} = c(1+z)(\Delta\nu_{obs}/\nu_\circ)$.}
\tablenotetext{c}{1.4 GHz continuum fluxes are courtesy of the NRAO VLA Sky Survey 
\citep{con98}.}
\end{deluxetable}

\end{document}